\documentclass[11pt]{article}

\usepackage[breakable]{tcolorbox}
\usepackage{appendix}
\usepackage{iftex}
\usepackage[T1]{fontenc}
\usepackage{comment}
\usepackage{caption}
\usepackage{subcaption}
\usepackage{graphicx}
\usepackage{xcolor, soul}
\sethlcolor{white}
\setulcolor{green}
\setstcolor{red}
\usepackage{float}
\usepackage{xcolor} % Allow colors to be defined
\usepackage{enumerate} % Needed for markdown enumerations to work
\usepackage{geometry} % Used to adjust the document margins
\usepackage{amsmath} % Equations
\usepackage{amssymb} % Equations
\usepackage{textcomp} % defines textquotesingle
\AtBeginDocument{%
}
\usepackage{upquote} % Upright quotes for verbatim code
\usepackage{eurosym} % defines \euro
\usepackage{fancyvrb} % verbatim replacement that allows latex
\usepackage{grffile} % extends the file name processing of package graphics 
\makeatletter % fix for old versions of grffile with XeLaTeX
\@ifpackagelater{grffile}{2019/11/01}
{
}
{
  \def\Gread@@xetex#1{%
    \IfFileExists{"\Gin@base".bb}%
    {\Gread@eps{\Gin@base.bb}}%
    {\Gread@@xetex@aux#1}%
  }
}
\makeatother
\usepackage[Export]{adjustbox} % Used to constrain images to a maximum size
\adjustboxset{max size={0.9\linewidth}{0.9\paperheight}}

\usepackage{hyperref}
\usepackage{titling}
\usepackage{longtable} % longtable support required by pandoc >1.10
\usepackage{booktabs}  % table support for pandoc > 1.12.2
\usepackage[inline]{enumitem} % IRkernel/repr support (it uses the enumerate* environment)
\usepackage[normalem]{ulem} % ulem is needed to support strikethroughs (\sout)
\usepackage{mathrsfs}
\usepackage[style=apa, citestyle=apa, backend=biber, uniquename=false, mincitenames=1, maxcitenames=2, citetracker, maxbibnames=99, doi=false, isbn=false, url=false, eprint=false]{biblatex}
\usepackage{amsmath}
\usepackage[ruled,vlined]{algorithm2e}
\usepackage[bottom,flushmargin]{footmisc}
\usepackage{authblk}
\usepackage{xcolor, soul}
\sethlcolor{green}
\usepackage{mathtools}

\usepackage{changepage}

\providecommand{\keywords}[1]
{
  \noindent
  \small	
  \textbf{Keywords:} #1
}

\DeclareMathOperator*{\argmin}{arg\,min}

\definecolor{urlcolor}{rgb}{0,.145,.698}
\definecolor{linkcolor}{rgb}{.71,0.21,0.01}
\definecolor{citecolor}{rgb}{.12,.54,.11}

\let\Oldtex\TeX
\let\Oldlatex\LaTeX
\renewcommand{\TeX}{\textrm{\Oldtex}}
\renewcommand{\LaTeX}{\textrm{\Oldlatex}}
\makeatletter
\AtEveryCitekey{%
  \ifciteseen
    {}
    {\defcounter{maxnames}{6}}%
}
\makeatother
\title{\textbf{Sample Fit Reliability\thanks{A previous version of this paper was presented at the University of Texas at Austin. We thank participants for useful comments and suggestions. We also thank Mirjam B\"achli, Daniel Goller, Sandro Heiniger, Maximilian Hofer, Michael Knaus and Jana Mareckova for their valuable feedback on previous versions of the manuscript as well as Hugo Bodory and Nikolas Kuschnig for helpful discussions.}}\\}

\author{Gabriel Okasa\thanks{\noindent Email: \href{mailto:okasa.gabriel@gmail.com}{gabriel.okasa@epfl.ch}} }
\author{Kenneth A. Younge\thanks{\noindent Email: \href{mailto:kenyounge@gmail.com}{kenneth.younge@epfl.ch}}}
\affil{EPFL\\\vspace{0.1cm}The Swiss Federal Institute of Technology in Lausanne\\~\\}
\date{September 14, 2022}

\hypersetup{
  breaklinks=true,  % so long urls are correctly broken across lines
  colorlinks=true,
  urlcolor=urlcolor,
  linkcolor=linkcolor,
  citecolor=citecolor,
  pdftitle={Sample Fit Reliability},
  pdfauthor={Gabriel Okasa and Kenneth A. Younge},
  pdfsubject={Econometrics},
  pdfkeywords={reliability, robustness, sensitivity, re-sampling},
  pdfstartpage=1}
\begin{document}
    
\maketitle
%\vspace{-0.5cm}
\begin{adjustwidth}{0.9cm}{0.9cm}
\begin{abstract}
%\vspace{0.1cm}
\noindent
Researchers frequently test and improve \textit{model fit} by holding a sample constant and varying the model. We propose methods to test and improve \textit{sample fit} by holding a model constant and varying the sample. Much as the bootstrap is a well-known method to re-sample data and estimate the uncertainty of the fit of parameters in a model, we develop \textit{Sample Fit Reliability} (SFR) as a set of computational methods to re-sample data and estimate the reliability of the fit of observations in a sample. SFR uses \textit{Scoring} to assess the reliability of each observation in a sample, \textit{Annealing} to check the sensitivity of results to removing unreliable data, and \textit{Fitting} to re-weight observations for more robust analysis. We provide simulation evidence to demonstrate the advantages of using SFR, and we replicate three empirical studies with treatment effects to illustrate how SFR reveals new insights about each study.\\

\end{abstract}
\end{adjustwidth}

\thispagestyle{empty}

\pagebreak

\setcounter{page}{1}

\section{Introduction}\label{intro}

Improving the reliability of empirical results is a core concern for all applied econometrics \parencite{leamer1983, leamer1985}. Researchers routinely assess the credibility of causal identification \parencite{angrist2010}, the selection of confounders \parencite{wuthrich2020}, the functional form of models \parencite{athey2017}, the properties of estimators \parencite{busso2014}, and the model uncertainty \parencites{lehrer2017}{steel2020}. Less attention, however, is dedicated to assessing and improving the data that is used to represent a phenomenon \parencite{kuschnig2021}. In other words, researchers often test \textit{model fit} (by holding data constant and varying the model), but rarely test \textit{sample fit} (by holding the model constant and varying the data).

Accepting data ``as-is'' can lead to studies that over-fit to noise \parencite{hastie2009} and fail to replicate to new data \parencites{hubbard1996empirical}. Unfortunately, some have concluded that empirical research is not to be trusted \parencites[e.g.,][]{ioannidis2005}. In comparison, fields such as data science and machine learning deal with bad data (including errors in measurement, labeling, data entry, disambiguation, data conversion, record merging, censoring, and so forth) with computational methods that can test the sensitivity and rubustify the results with respect to the underlying data \parencites[see e.g.,][]{fischler1981}{liu2008}{zheng2016}{schouten2018}{cohen2020}. Much as the ``credibility revolution'' in applied econometrics  \parencite{angrist2010} shifted attention to the importance of research design, we argue that more attention in the field should be focused on data reliability. We believe computational methods offer promising tools and solutions to that end.

In this paper, we provide a data-focused perspective for assessing sample fit in econometric models. In particular, we suggest a re-sampling procedure to estimate the reliability of the data, check the sensitivity of the results, and improve the robustness of the analysis. Accordingly, we propose \textit{Sample Fit Reliability} (SFR) as a computational approach with three aspects: \textit{Scoring}, to estimate a reliability score for every observation in a sample based on the expected estimation loss over sub-samples; \textit{Annealing}, to test the sensitivity of results to the sequential removal of unreliable data; and \textit{Fitting}, to estimate a weighted regression that adjusts for the reliability of the sample. To demonstrate the utility of SFR, we provide an example based on real-world data, test the performance of SFR using synthetic data in simulations, and replicate renowned field experiments in economics. We find that, in each case, SFR methods improve understanding of sample fit and reveal valuable insights about the reliability of results.

Our contribution is threefold. First, we put forward the concept of sample fit as a complement to the widely used notion of model fit. Second, we combine ideas from classical robust statistics with modern machine learning to develop SFR. Third, we introduce a graphical approach to sensitivity analysis that uses annealing based on reliability scores to provide insights about the stability and heterogeneity of effects. Finally, we provide an implementation of SFR in a Python package called \href{https://pypi.org/project/samplefit/}{samplefit} that is available on the PyPI repository for future use by applied researchers.

\section{Related Work}\label{sec:lit}

Concerns about unreliable data traditionally have been addressed via robust estimation with a long tradition in statistics \parencite{rousseeuw2005}. In particular, M-estimation methods such as Huber regression \parencite{huber1964} acknowledge the instability of OLS in the presence of outliers. M-estimation modifies the loss function and incorporates weighting based on residuals \parencites{fox2002}{yu2017}. Another stream of literature aims to identify sets of ``influential observations'' \parencite{cook1979}. Influential observations are data points that strongly impact quantities of interest, including model parameters, when those observations are omitted \parencite{belsley2005}. The measurement of influential data points generally relies on leverage and the magnitude of residuals \parencite{chatterjee1986}, or a similar assessment of influence \parencites{hampel1974}{hampel2011}.

Researchers also use re-sampling methods to test, diagnose, and improve results by perturbing the underlying data \parencite{yu2013}. Re-sampling methods include techniques such as the bootstrap, the jackknife, cross-validation, and other forms of sub-sampling \parencites{efron1982}{efron1994}. Bootstrapping, for example, is used in econometrics to quantify sampling uncertainty for estimated parameters \parencites{diciccio1996}{horowitz2001}{mackinnon2006}; bootstrap aggregation (``bagging'') is used in machine learning to reduce variance and improve predictions \parencites{breiman1996}{buhlmann2002}; bootstrapping and sub-sampling are used together in ensemble methods such as random forests \parencite{breiman2001} and causal forests \parencites{lechner2018}{wager2018}{wager2019}; and sub-sampling is used in computer vision to separate inliers vs. outliers \parencite{raguram2008}.

Additionally, \textcite{grandvalet2001, grandvalet2004} points out that bagging stabilizes estimation in the presence of a few influential points by indirectly down-weighting those points, similarly to robust regression methods. In computer vision, re-sampling techniques such as the Random Sample Consensus (RANSAC) algorithm \parencite{fischler1981} drop outliers and influential points, instead of re-weighting. In particular, RANSAC repeatedly draws random sub-samples of the data and assesses the fit of data points excluded from the sub-sample. Random sub-sampling is repeated until the number of inliers reaches a given criteria, and a final model is fitted on a ``consensus'' sub-sample of inliers. Random sub-sampling is also used in the optimization of the Least Median of Squares \parencite{rousseeuw1984b}, S-estimators \parencites{rousseeuw1984}{koller2017}, and forward search regression models \parencites{atkinson2007}{atkinson2010}.

Our work relates to several recent studies that address aspects of sample fit. Works by \textcite{broderick2020}, \textcite{kuschnig2021} and \textcite{moitra2022} identify data points within a sample that distort the estimation of model parameters when they are removed from an analysis in an adversarial manner. For example, \textcite{broderick2020} develop a computationally-efficient method to approximately identify the maximum influence perturbation (AMIP). As such, AMIP provides information about the worst-case scenario -- i.e., situations where the estimation of a parameter can be distorted by removing a minimum proportion of the sample. \textcite{kuschnig2021} further adapt perturbation testing to cases where jointly influential data points are a concern \parencite[i.e. masking, see ][]{rousseeuw1990}, and \textcite{moitra2022} provide theoretical guarantees for low dimensional cases.

\section{Sample Fit Reliability}\label{sec:framework}

We develop \textit{Sample Fit Reliability} as a computational approach that builds on insights from both classical robust statistics and modern machine learning. We begin by focusing on residuals as the fundamental diagnostic tool for regression. However, as pointed out by \textcite{belsley2005}, OLS residuals tend to underestimate the true error of outliers due to the squared error criterion. Therefore, we adopt a re-sampling procedure similar to RANSAC and compute the absolute loss based on residuals from many minimal random sub-samples; doing so reduces dependency on the results from any one sub-sample, and helps to alleviate the underestimation of the errors of outliers. Furthermore, re-sampling explores a wide range of possible model fits and stabilizes the estimation of loss through averaging. Thus, in contrast to RANSAC, we do not automatically search for a ``best'' sub-sample (e.g., a sub-sample free of outliers). Instead, we efficiently use all of the information from each sub-sample to estimate the \textit{expected} loss for each observation. We therein avoid arbitrary specifications of error thresholds. Finally, we define reliability scores to be reversely related to the expected losses.

Below, we define the target quantity of interest, the estimation algorithm, a new approach to sensitivity analysis, and a robust re-weighting procedure for estimation. We focus on the parametric family of models, and in particular on the linear regression models, of type:

$$Y_i = X_i'\beta + \epsilon_i$$

where $Y_i$ is the outcome, $X_i$ is a vector of covariates, $\beta$ is a vector of unknown coefficients, and $\epsilon_i$ is the error term.

There are several reasons why we focus on linear models. First, linear models are still the dominant models for causal inference \parencites{angrist2008}{imbens2015}, despite recent developments in causal machine learning \parencite[see][for an overview]{athey2019}. For example, selection-on-observables via covariate adjustment, instrumental variables via two-stage least squares, and difference-in-differences via fixed effects are typically all estimated in linear models. As such, current econometric research focuses on addressing problems with estimating treatment effects via OLS in the presence of effect heterogeneity \parencites{sloczynski2022}{pinkham2022}. Similarly, our research addresses problems with estimating effects -- including treatment effects -- in the presence of unreliable data. Second, and in comparison to flexible machine learning models, linear models provide a high degree of interpretability via marginal effects. Interpretability is valuable for communicating results to policy makers and is often required in regulated contexts. Our work therefore supports the advancement of simpler, more-interpretable models by restricting and/or re-weighting the sample according to the reliability of the data. Third, data in the social sciences is often scarce, and collecting additional data is generally expensive. As such, often there is insufficient data for highly-flexible, non-parametric models \parencite{racine2007}.

To address possible data issues, we are interested in estimating the reliability of each data point in a sample, conditional on a given model specification. We define our quantity of interest as follows. Suppose that we have access to $i=1,...,N$ training data $Z_i=(X_i, Y_i)$ and an input $x$. Consider an OLS estimator of the conditional mean $M(x;Z_i)=\mathbb{E}[Y_i \mid X_i=x]= X_i'\beta$. We are interested in the estimation error $\gamma(x)$ given a loss function as $\hat{\gamma}(x)=\mathcal{L}\big(Y_i, \hat{M}(x;Z_i)\big)$. With re-sampling we aim to stabilize the estimator of the loss as:

$$\hat{\Gamma}^{\infty}(x) = \mathbb{E}^*\big[\mathcal{L}\big(Y_{i}^*, M(x;Z_{i}^*)\big)\big]$$

where $Z_{i}^*$ are sub-samples drawn without replacement and the expectation $\mathbb{E}^*$ is taken with respect to the re-sampling measure. To evaluate the expectation, we rely on sub-sampling aggregation and form the Monte Carlo estimator as follows:

$$\hat{\Gamma}^{S}(x) = \frac{1}{S}\sum^S_{s=1} \gamma_s^*(x) \text{ where } \gamma_s^*(x)=\mathcal{L}\big(Y_{i,s}^*, M(x;Z_{i,s}^*)\big)$$

with $Z_{i,s}^*$ being the $s$-th sub-sample. Thus, for $S \rightarrow \infty$ we approximate $\hat{\Gamma}^{\infty}(x)$. Finally, we define the reliability score as a reverse transformation of the expected loss as follows:

$$\hat{\psi}(x) = \frac{\hat{\Gamma}^{S}(x) - \max\{\hat{\Gamma}^{S}(X_1,...,X_N)\}}{\min\{\hat{\Gamma}^{S}(X_1,...,X_N)\} - \max\{\hat{\Gamma}^{S}(X_1,...,X_N)\}}$$

to ensure boundedness of reliability scores as $0 \geq \hat{\psi}(x) \geq 1$. A reliability score of 0 corresponds to the largest expected loss. A reliability score of 1 corresponds to the smallest expected loss.

\subsection{Scoring}\label{sec:algorithm}

The SFR algorithm for \textit{Scoring} the reliability of each data point in a sample is defined below. Consider an $iid$ random sample of size $i=1,...,N$ with random variables consisting of features $X_i$ of dimension $p$ and outcomes $Y_i$. SFR draws a random sub-sample $(Z^*_{i,s})_{i\in A_s}$ of size $\eta=p+1$ from the original data $Z_i=(X_i, Y_i)$ without replacement,\footnote{Such sub-sampling is close to the elemental sets sampling, where sub-sample size is exactly equal to $p$ \parencite[see e.g.][]{olive2007}. In practice, we find the almost elemental sets of size $p+1$ to have more stable performance.} where $A_s$ is the set of indices corresponding to the $s$-th sub-sample for $s=1,...,S$ sub-sampling iterations. For each sub-sample $(Z^*_{i,s})_{i\in A_s}$, SFR estimates the model $M_s\big(x;(Z^*_{i,s})_{i\in A_s}\big)$ via OLS and evaluates the corresponding loss, i.e. $\hat{\gamma}_{s}(X_{(-i)})$, for each out-of-bag data point $(Z_{i,s})_{i\notin A_s}=(X_{(-i)}, Y_{(-i)})$, given a loss function $\mathcal{L}\big(Y_{(-i),s}^*, \hat{M}(X_{(-i)};Z_{i,s}^*)\big)$,\footnote{We use an absolute loss similarly to RANSAC \parencite{fischler1981}.} where $\hat{M}_s(X_{(-i)};Z_{i,s}^*)$ is the model prediction for an out-of-bag observation $X_{(-i)}$. The losses for an individual data point $i$ are subsequently averaged over the $S$ sub-sampling iterations in which the data point $i$ was not used for estimation, i.e. out-of-bag losses, as $\Gamma^S(X_i) = \frac{1}{S}\sum^S_{s=1} \gamma_s(X_{(-i),s})$.\footnote{The number of possible sub-samples increases with order $N^{\eta}$ and it becomes prohibitive to explore all of them. Yet, as pointed out by \textcite{koller2017} in the case of S-estimators, in practice setting S=1000 is sufficient.} A reliability score $\psi_i$ for data point $i$ is then reversely related to the average loss, such that the largest (smallest) average loss corresponds to a reliability score of zero (one). The reliability scores are effectively bounded between 0 and 1 through $\hat{\psi}_i = \frac{\hat{\Gamma}^{S}(X_i) - \max\{\hat{\Gamma}^{S}(X_1,...,X_N)\}}{\min\{\hat{\Gamma}^{S}(X_1,...,X_N)\} - \max\{\hat{\Gamma}^{S}(X_1,...,X_N)\}}$. Intuitively, if an observation exhibits a large (small) loss throughout many random sub-samples, relative to the other observations in a sample, we define such an observation as having a low (high) reliability.\\

\begin{algorithm}[H]
\SetAlgoLined
\KwIn{Data: $Z_i$, Iterations: $S$, Sub-sample size: $\eta$, Model: $M(x)$, Loss: $\mathcal{L}(y,\hat{M}(x))$}
\KwOut{Reliability Scores: $\hat{\psi}_i$}
\Begin{
 \For{$s=1$ \KwTo $S$}{
 \phantom{...}Sample $(Z^*_{i,s})_{i\in A_s}$ of size $\eta$ from $Z_i=(X_i,Y_i)$ without replacement\;
 \phantom{...}Estimate the model $M\big(x;(Z^*_{i,s})_{i\in A_s}\big)=\mathbb{E}[Y_{i,s} \mid X_{i,s}=x]$ via OLS\;
 \phantom{...}Evaluate loss $\hat{\gamma}_s(X_{(-i),s})=L\big(Y_{(-i),s}, \hat{M}(X_{(-i)};(Z^*_{i,s})_{i\in A_s})\big)$ out-of-bag\;
 }
 Average over the losses as $\hat{\Gamma}^S(X_i) = \frac{1}{S}\sum^S_{s=1} \hat{\gamma}_s(X_{(-i),s})$ out-of-bag\;
 Define reliability scores as $\hat{\psi}_i = \frac{\hat{\Gamma}^{S}(X_i) - \max\{\hat{\Gamma}^{S}(X_1,...,X_N)\}}{\min\{\hat{\Gamma}^{S}(X_1,...,X_N)\} - \max\{\hat{\Gamma}^{S}(X_1,...,X_N)\}}$ for $i=1,...,N$\;
}
\caption{\textsc{Sample Fit Reliability (SFR)}}
\label{algo:SFR}
\end{algorithm}
~\\

As can be seen in Algorithm \ref{algo:SFR}, the main output of SFR is a full complement of reliability scores $\hat{\psi}_{i}$ across all observations in the sample. The scores postulate a continuous, point-wise measure of data reliability based on the reverse relationship to the average losses, i.e. higher average losses imply lower reliability scores. Then, reliability scores can be inspected and investigated to generate insights about sources of error (e.g., a corrupt process for capturing data, censoring, measurement error, or a secondary data generating process).

\begin{figure}[H]
    \centering
    \includegraphics[scale=0.7]{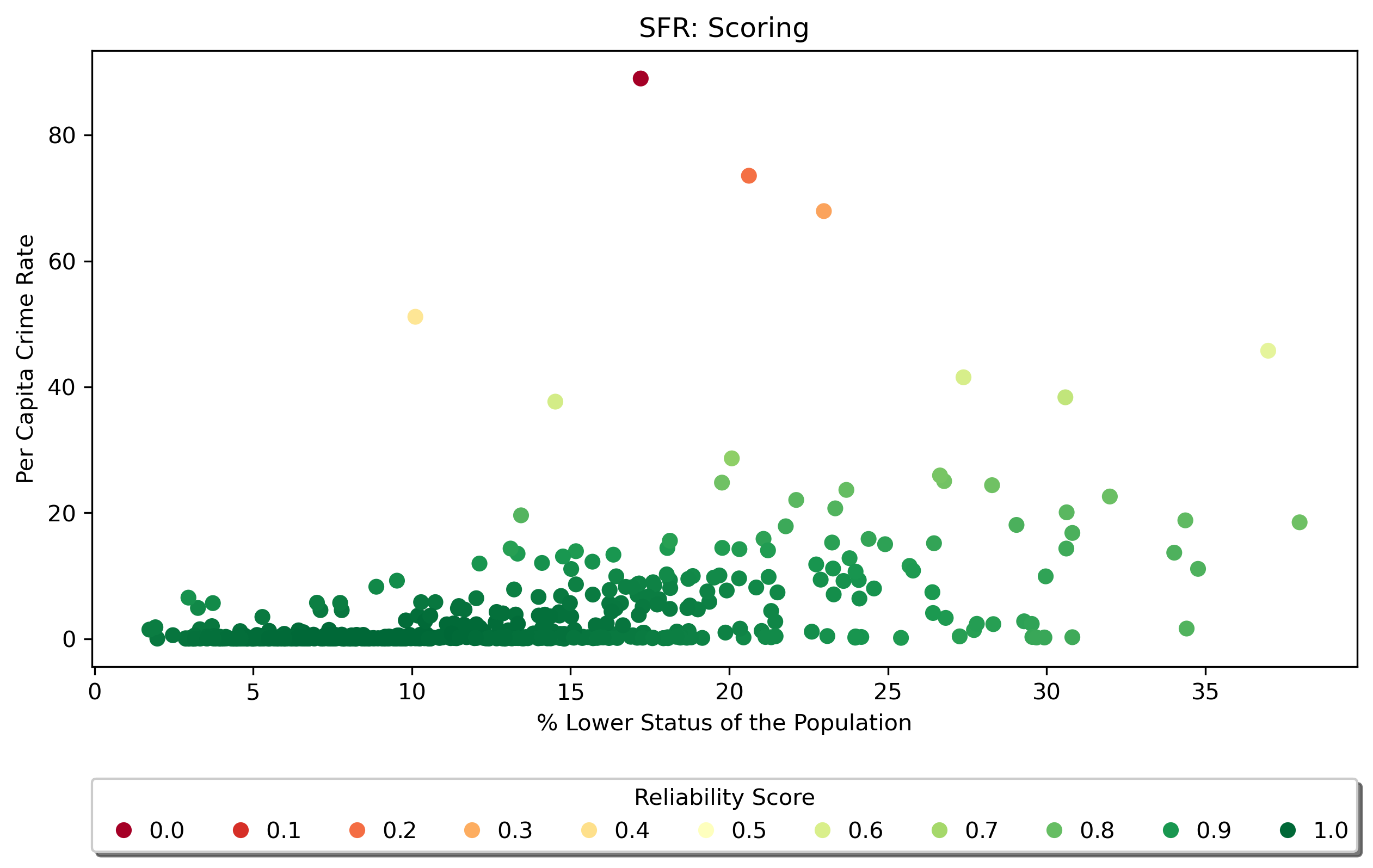}
    \caption{Scoring - Boston Housing Data}
    \label{fig:reliability-scores}
\end{figure}

The practical implications of scoring reliability are illustrated in Figure \ref{fig:reliability-scores}, based on the popular Boston housing dataset \parencites{harrison1978}{belsley2005}.\footnote{Descriptive statistics are provided in Appendix \ref{app:boston}.} We consider the relationship between the per capita crime rate and the percentage of the population that is considered to be lower status in a neighborhood (e.g. lack of education, labor work, etc.). The scatterplot of the data suggests a slightly positive relationship, distorted by a few neighborhoods that have an unusually high level of crime, but a moderate level of lower status individuals. Figure \ref{fig:reliability-scores} shows that SFR naturally identifies unusual data points and assigns the reliability scores accordingly. In particular, we see that reliability scores are lower for observations that are further from the center of the distribution and/or reinforced by less evidence. Such diagnostics of sample fit can help to draw attention to conditions where the model does not generalize well.

\subsection{Annealing}\label{annealing}

\begin{figure}[H]
    \centering
    \includegraphics[scale=0.7]{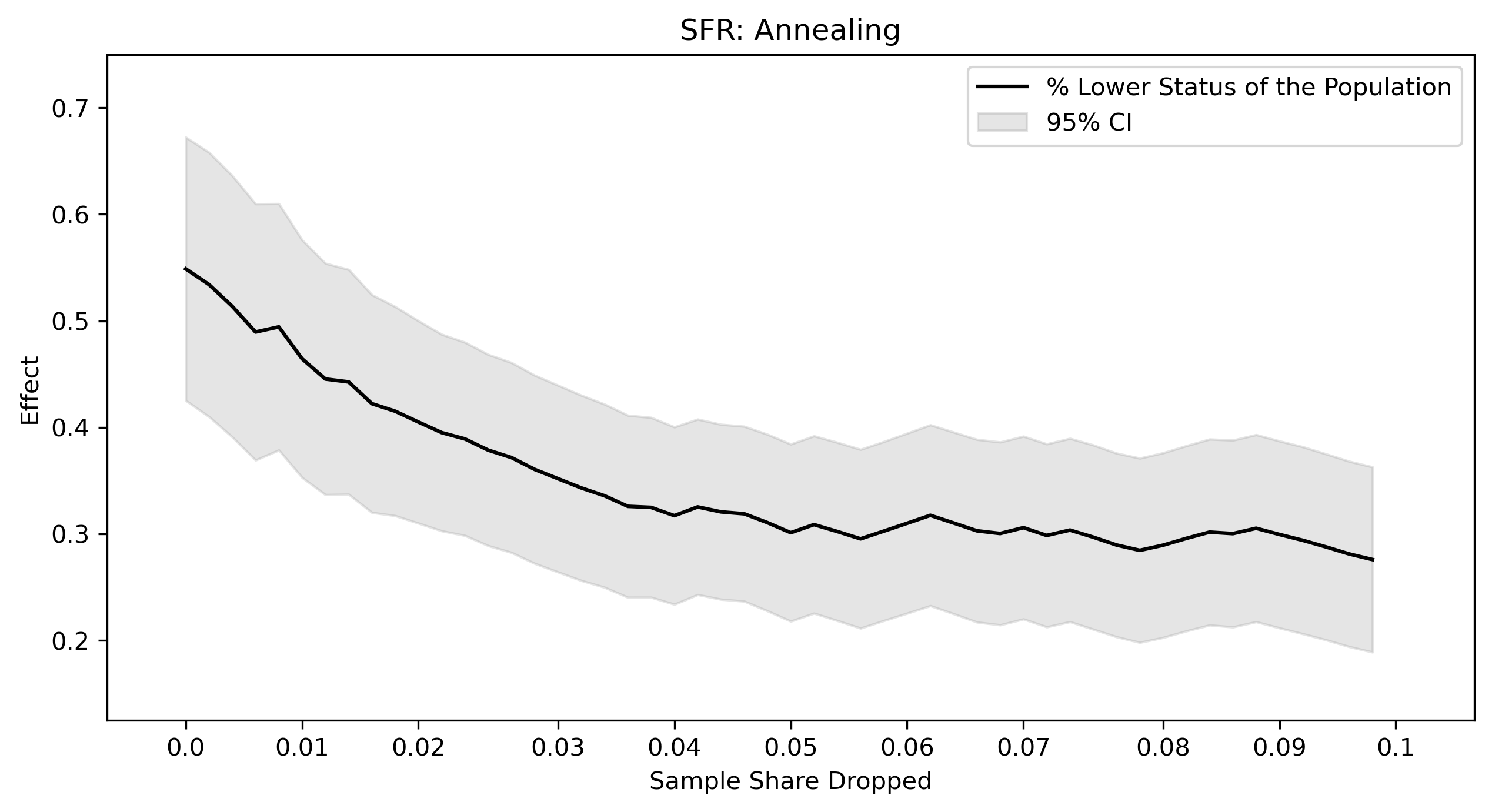}
    \caption{Annealing - Boston Housing Data}
    \label{fig:annealing-sensitivity}
\end{figure}

Researchers can sort reliability scores into increasing order (i.e. $\hat{\psi}_{(1)} \leq \hat{\psi}_{(2)} \leq \cdots \leq \hat{\psi}_{(N)}$) for a continuous measure of reliability across the sample. Sorted reliability scores $\hat{\psi}_{(i)}$ then can be used to assess the sensitivity of sample fit by re-estimating the model, while incrementally dropping unreliable data points from the sample. Accordingly, the estimation of a quantity of interest can be visualized along the annealing path.\footnote{Previously, \textcite{riani2014} advocated for graphical monitoring of robustness via related concept of forward plots \parencite[see][for details]{atkinson2007}.} More specifically, \textit{Annealing} drops a specified share of the most unreliable observations in a sequential manner to capture the sensitivity of parameter estimates to unreliable data.\footnote{The appropriate level of annealing will depend on the particular dataset and the amount of evidence required by the researcher; however, we have found that a $10\%$ annealing share is sufficient for many practical applications.} To reflect the uncertainty of estimated parameters, we use bootstrap approximation to take the two steps of the estimation procedure (both the estimation of reliability scores and the subsequent annealing) into account for inference. Here we follow \textcite{bodory2020} who show favourable performance of bootstrapping compared to asymptotic approximations for matching and weighting estimators that use estimated propensity scores and trim the sample accordingly \parencites(see for example)(for a discussion of trimming based on propensity scores)[][or]{crump2009}{lechner2019}. We illustrate the annealing based on the example from the Boston housing dataset in Figure \ref{fig:annealing-sensitivity}.

In Figure \ref{fig:annealing-sensitivity}, we see that the estimated effect decreases as we remove the most unreliable data points, and the estimated effect stabilizes once approximately $5\%$ of the sample has been annealed. After $5\%$, further annealing of the data does not substantially change the point estimate. The annealing graph shows that the original effect estimate might not provide an appropriate description of an underlying relationship. We also note that restricting the sample through annealing essentially estimates the effect for different sub-populations, as defined by reliability scores. Therefore, it is recommended to check the covariate balance between the restricted and unrestricted sample in order to prevent over-generalization of the results \parencite{diamond2013}. The annealing graph also draws the attention of researcher to the data points in a sample that may merit further inspection. For example, researchers might investigate the source of unreliability (i.e., measurement errors, different sub-populations, or a mis-specified model), and adapt accordingly (e.g., collect new data or specify new model). 

\subsection{Fitting}\label{fit}

In the sections above, we demonstrate how to use SFR to score the reliability of data and then perform a graphical, post-estimation sensitivity analysis of sample fit via annealing. In some cases, researchers may then identify problems with data that justify dropping a portion of the sample, and researchers may then continue their analysis with the cleaned sub-sample. In the following section, however, we test how well a regression weighted by reliability scores can make an automatic analysis more robust.

SFR can be used for \textit{Fitting} a given model $M(x;Z_i)$ to the sample weighted by reliability scores $\hat{\psi}_i$, or their transformations $g(\hat{\psi}_i)$, wherein data points with higher reliability scores are up-weighted, and conversely data points with lower reliability scores are down-weighted. Similar forms of weighting are ubiquitous in most robust regression analysis \parencite[see for example][]{gervini2002}. We suggest a squared transformation of reliability scores, i.e. $g(\hat{\psi}_i)=\hat{\psi}_i^2$, that further accentuates the down-weighting of low reliable observations.\footnote{See for example \textcite{nie2021} for a similar squared weighting scheme to down-weight influence of extreme propensity scores in a non-parametric treatment effect estimation.}

We thus apply the weighted least squares estimator as follows:

$$\hat{\beta}=\argmin_{\beta}\sum_{i=1}^N g(\hat{\psi}_i)\cdot\big(Y_i - X'_i\beta\big)^2$$

to estimate the parameters of the model $M(x;Z_i)=\mathbb{E}[Y_i \mid X_i=x]=X'_i\beta$.\footnote{Given our definition of reliability, with the most unreliable data point having a score of zero, weighting will effectively discard one data point. The loss of information, however, is of order $(1/N)$ and will vanish asymptotically.} We apply non-parametric pairs bootstrap \parencite{diciccio1996} to approximate standard errors due to the two-step estimation procedure described above (i.e., SFR first estimates reliability scores and then estimates a weighted regression using the estimated scores). \textcite{diciccio2019} establish the consistency of the bootstrap approximation for the sampling distribution of the weighted least squares estimator and further show its better coverage in finite samples in comparison to asymptotic approximations.

\begin{table}[H]
\small
\centering
\caption{Fitting - Boston Housing Data}
\label{tab:weighted-fit}
\begin{tabular}{lrrrrrr}
\toprule
{} &  Coef. &  Std.Err. &       t &  P>|t| &  [0.025 &  0.975] \\
\midrule
OLS & 0.5488 &    0.0478 & 11.4907 & 0.0000 &  0.4550 &  0.6426 \\
SFR & 0.4315 &    0.0427 & 10.1120 & 0.0000 &  0.3476 &  0.5153 \\
\bottomrule
\multicolumn{7}{l}{\footnotesize{\textit{Note:} Coefficients for the slope of the \% lower status of the population.}}
\end{tabular}
\end{table}

Table \ref{tab:weighted-fit} presents results of a reliability-weighted fit for the Boston housing dataset examined earlier. We see that the reliability-weighted fit supports a better description of the underlying relationship by effectively down-weighting the influence of unreliable data points. Doing so results in a lower marginal effect, in line with conclusions drawn from the annealing exercise.

\section{Simulations}\label{sec:simulations}

In this section, we investigate the performance of SFR with simulated data in four stylized scenarios, with a sample size of $N=1000$ and an outlier share of $\alpha = 5\%$. In Appendix \ref{app:sim}, we vary $N$ and $\alpha$ and obtain qualitatively similar results.

\textbf{Scenario 1: No Outliers.} In a baseline scenario, we simulate $i=1,...,N$ data points from a linear model $Y_i= X_i'\beta+\epsilon_i$, where $X_i \sim \mathcal{N}(0,1)$, $\beta=1$, and $\epsilon_i \sim \mathcal{N}(0,2/3)$. OLS is the best linear unbiased estimator in this scenario.

\textbf{Scenario 2: Randomly Distributed Outliers.} In the second scenario, we simulate data in the same manner as in the first scenario, but include a set $\mathcal{B}$ of size $\alpha\times N$ consisting of randomly distributed outliers via simulating data points $X_i \sim \mathcal{N}(0,2)$ with a corresponding outcome $Y_i \sim \mathcal{N}(0,2)$ for $i \in \mathcal{B}$. In this case, one would expect random distortions of the OLS estimator.

\textbf{Scenario 3: Densely Distributed Outliers with Leverage.} In the third scenario, we simulate the data again as in the first scenario, but in this case include a set $\mathcal{B}$ of size $\alpha\times N$ of densely distributed outliers with high leverage by simulating data points $X_i \sim \mathcal{N}(2,1/2)$ with a corresponding outcome $Y_i \sim \mathcal{N}(-1,2/3)$ for $i \in \mathcal{B}$. Scenario 3 is comparable to creating an influential set as in \textcite{broderick2020}, including the masking phenomenon as in \textcite{kuschnig2021}. The OLS estimator is expected to have a negative bias that would attenuate the main effect.

\textbf{Scenario 4: Clusters of Densely Distributed Outliers with Leverage.} In the fourth scenario, we simulate data in the same manner as in Scenario 3, except that we split the set $\mathcal{B}=\mathcal{B}_1 \cup \mathcal{B}_2$ into two equal halves and simulate data points $X_i \sim \mathcal{N}(2,1/2)$ with a corresponding outcome $Y_i \sim \mathcal{N}(-1,2/3)$ for $i \in \mathcal{B}_1$ and data points $X_i \sim \mathcal{N}(-2,1/2)$ with a corresponding outcome $Y_i \sim \mathcal{N}(1,2/3)$ for $i \in \mathcal{B}_2$, respectively. Hence, we create two distinct clusters of outliers with leverage that are expected to bias the OLS estimator in a similar way as in Scenario 3.

\subsection{Scoring}

\begin{figure}[H]
    \centering
    \includegraphics[scale=1.05]{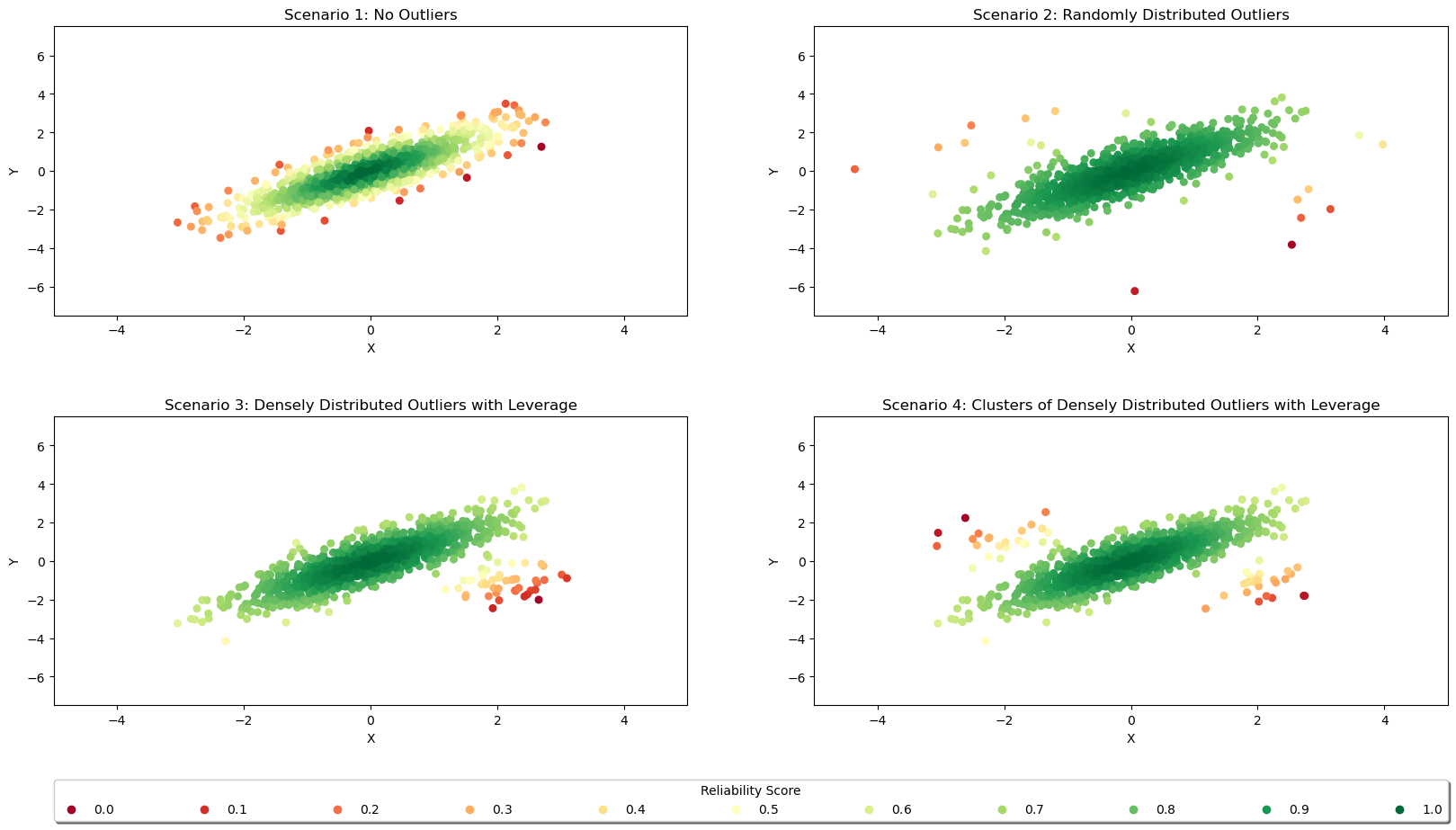}
    \caption{Scoring - Comparison of Simulation Scenarios}
    \label{fig:scores-sim}
\end{figure}

In Figure \ref{fig:scores-sim}, we present graphical results from estimating reliability scores based on the synthetic data described above. In Scenarios 2 through 4, we see that SFR accurately scores the reliability of observations corresponding to both clustered and randomly distributed outliers. Notice that reliability scores are both distance \textit{and} density sensitive -- reliability scoring reflects the distance to the regression line and the amount of information around each point. As such, leverage points that comply with the imposed regression line are scored with lower reliability, even though they would not distort the estimation. Moreover, leverage points that stand off the imposed regression line and do distort the estimation, are scored with even lower reliability. As a consequence, even in Scenario 1 without outliers, SFR assigns lower reliability scores to data points with low density surrounding the imposed regression line such that the estimation would not be distorted. Hence, the simulation provides evidence that SFR scoring appropriately evaluates the reliability of the data.

\subsection{Annealing}

\begin{figure}[H]
    \centering
    \begin{subfigure}{0.495\textwidth}
    \centering
    \includegraphics[scale=1.2]{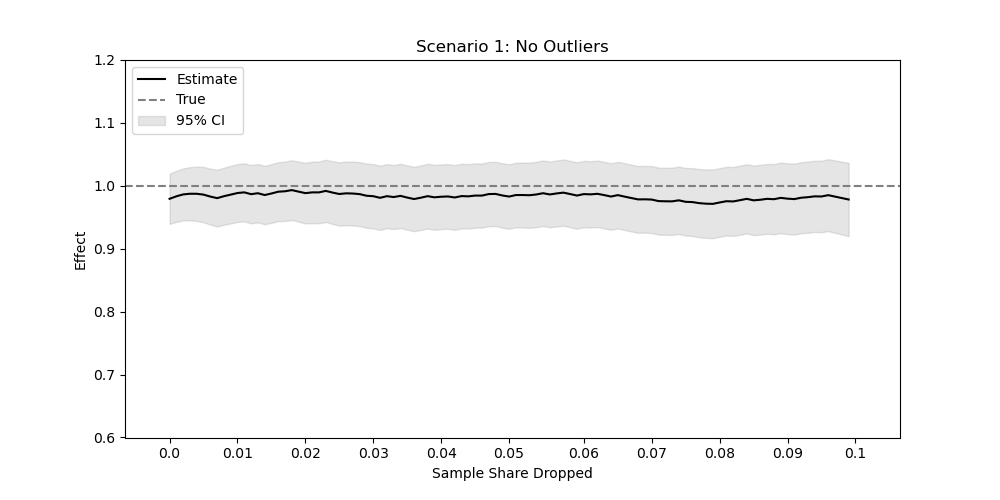}
    \label{fig:scores-a}
    \end{subfigure}
    \begin{subfigure}{0.495\textwidth}
    \centering
    \includegraphics[scale=1.2]{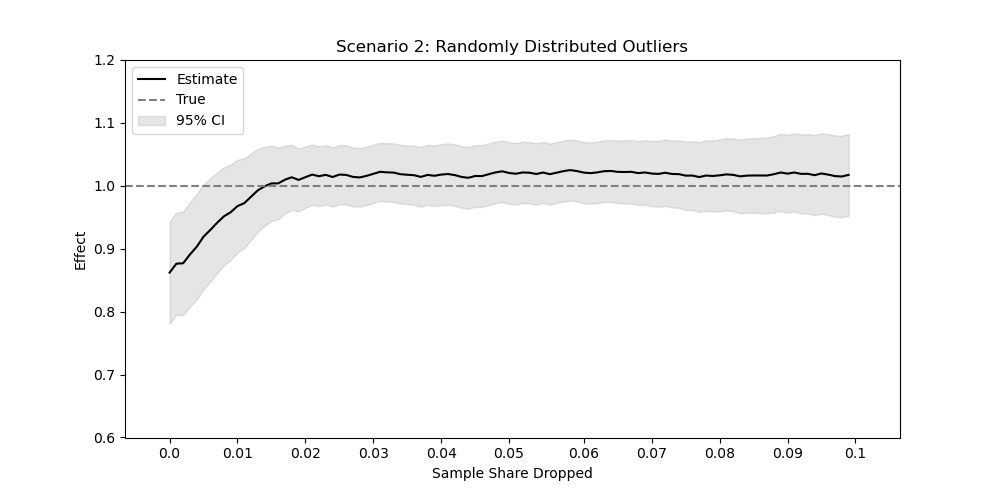}
    \label{fig:scores-b}
    \end{subfigure}
    \vspace{-0.5cm}
    \begin{subfigure}{0.495\textwidth}
    \centering
    \includegraphics[scale=1.2]{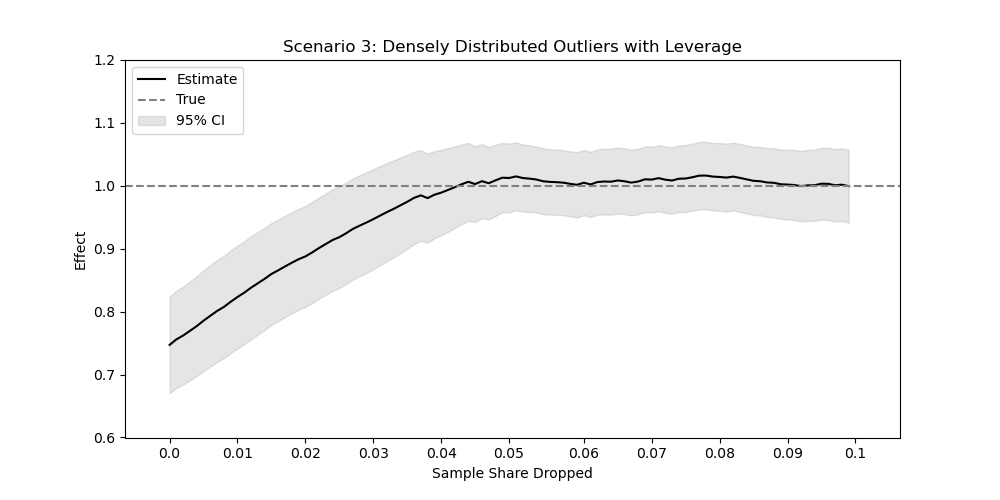}
    \label{fig:scores-c}
    \end{subfigure}
    \begin{subfigure}{0.495\textwidth}
    \centering
    \includegraphics[scale=1.2]{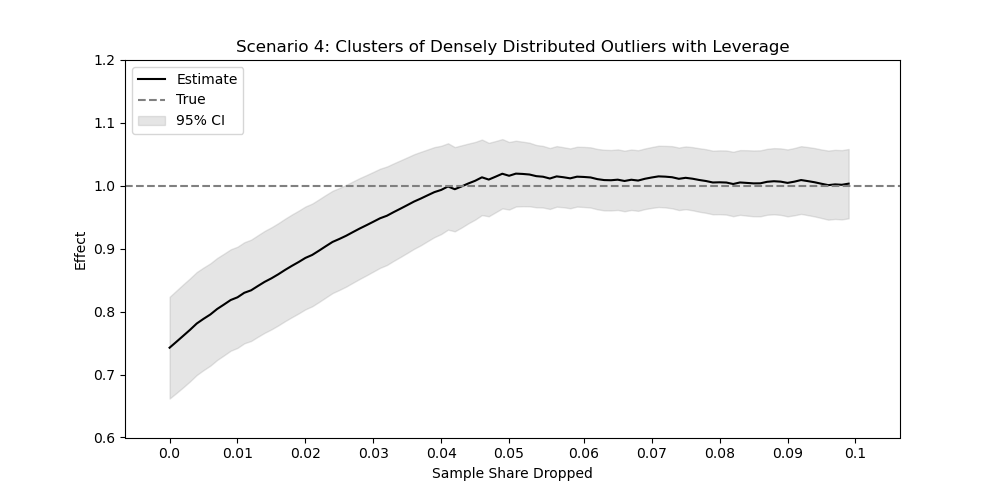}
    \label{fig:scores-d}
    \end{subfigure}
    \caption{Annealing - Comparison of Simulation Scenarios}
    \label{fig:annealing}
\end{figure}

To go beyond an assessment of the reliability of individual data points, and to gain substantive insights as to how unreliable data can affect the estimation of parameters for a given model, we apply the annealing procedure described above to assess the sensitivity of effects to the exclusion of unreliable data. We present graphical results from the annealing procedure for the synthetic datasets in Figure \ref{fig:annealing}. For Scenario 1 (no outliers), annealing has no effect; for Scenarios 2, 3 and 4 (with outliers), annealing converges to the true underlying effect $\beta=1$ after removing the first $5\%$ of unreliable data points (i.e., the same $5\%$ included in set $\mathcal{B}$ during simulation of the synthetic data).\footnote{Note, that the annealing converges faster to the true effect in Scenario 2 as some of the randomly distributed outliers overlay with the true underlying DGP.} We therefore conclude that annealing can help to detect unstable effects related to outliers, without distorting results that are free of outliers.

\subsection{Fitting}
Finally, we evaluate the performance of a weighted regression based on reliability scores for 1,000 replications of each scenario. Table \ref{tab:simeffect} reports the results for the mean squared error, the mean absolute bias, the standard deviation and the Jarque Bera statistic of normality, for the estimation of the effect (slope) parameter. In column 1 (no outliers), OLS estimation is theoretically the optimal choice. Nevertheless, SFR exhibits low bias and slightly lower efficiency, while being normally distributed as well. In column 2 (randomly distributed outliers), the bias for OLS is considerably larger, and so is its variance. In contrast, SFR performs substantially better than OLS, both in terms of bias and MSE. In column 3 (single cluster of densely distributed outliers with high leverage), OLS is noticeably biased, and is again outperformed by SFR. Finally, in column 4 (two clusters of densely distributed outliers with high leverage), OLS again has high bias, while SFR correctly addresses the presence of outliers and returns bias and MSE that is an order of magnitude lower than that of OLS, while being comparably efficient.

Appendix \ref{app:sim} reports additional results for comparisons with the Huber and RANSAC estimators. SFR is more robust than both Huber and RANSAC in the simulations, especially against high leverage points. A known drawback of the Huber estimator is sensitivity to high leverage points \parencite{yu2017}, and the simulations demonstrate high variance for the RANSAC estimator, although RANSAC's performance is expected to be better in datasets with higher outlier contamination \parencite{raguram2008}. Across the simulations, we conclude that SFR compares favourably to both OLS and robust regression methods, when an automated analysis is preferred to the researcher-driven methods for scoring and annealing covered in previous sections.

\begin{table}[H]
\small
 \centering
 \caption{Fitting - Comparison of Simulation Scenarios}\label{tab:simeffect}
 \begin{tabular}{lcccc}
 \toprule
 & \multicolumn{4}{c}{\textit{\textbf{Simulation Scenario}}}\\
 \midrule
 & (1) & (2) & (3) & (4)\\
 \midrule
& \multicolumn{4}{c}{\textit{Mean Squared Error}}\\
 \midrule
OLS &    0.0005 &     0.0313 &     0.0674 &     0.0734 \\
SFR &    0.0006 &     0.0036 &     0.0091 &     0.0085 \\
\midrule
& \multicolumn{4}{c}{\textit{Mean Absolute Bias}}\\
\midrule
OLS &    0.0169 &     0.1714 &     0.2585 &     0.2696 \\
SFR &    0.0201 &     0.0536 &     0.0918 &     0.0886 \\
\midrule
& \multicolumn{4}{c}{\textit{Standard Deviation}}\\
\midrule
OLS &    0.0213 &     0.0440 &     0.0243 &     0.0258 \\
SFR &    0.0253 &     0.0273 &     0.0258 &     0.0256 \\
\midrule
& \multicolumn{4}{c}{\textit{Jarque Bera Statistic}}\\
\midrule
OLS &    0.8285 &     0.0001 &     0.4811 &     0.5905 \\
SFR &    0.6744 &     0.1535 &     0.6346 &     0.2735 \\
\bottomrule
\multicolumn{5}{l}{\footnotesize \textit{Note:} $p$-values for Jarque Bera statistic reported.}
\end{tabular}
\end{table}

\section{Replications}\label{sec:replications}

Replication and reproducibility is an important aspect of all empirical research \parencites{anderson2008}{azoulay2015}{berry2017}{hoffler2017}. In the following section, we replicate three renowned field experiments from economics. In each case, the average treatment effect is identified via randomization and estimated by OLS in the following linear model:
\begin{equation}\label{eq:lm}
Y_i=\alpha + W_i\theta + \epsilon_i
\end{equation}

where $Y_i$ is the outcome, $\alpha$ is the intercept, $W_i$ denotes the binary treatment indicator, $\theta$ measures the corresponding average treatment effect (ATE) and $\epsilon_i$ is the error term.

In the first replication, we revisit the well-known field experiment in labor economics by \textcite{lalonde1986}. \textcite{lalonde1986} evaluates the impact of job training on future yearly earnings in US dollars based on a dataset of 445 individuals (henceforth the ``labor data''), where 185 were randomly assigned to join the National Supported Work (NSW) demonstration program in the US \parencite{dehejia1999}. Appendix \ref{app:training} reports descriptive statistics for the dataset.

In the second replication, we revisit the randomized control trial of \textcite{angelucci2015}. \textcite{angelucci2015} follow a vast literature on microcredit (for an overview, see \cite{banerjee2015} or \cite{meager2019}) and study the impact of access to microcredit on household welfare in Mexico. The sample includes data for more than 16,000 households (henceforth the ``microcredit data''), out of which approximately half received access to microcredit through door-to-door loan promotion. We follow \textcite{broderick2020} and \textcite{kuschnig2021} and re-evaluate the impacts of the microcredit provision on household profit in US dollars (PPP per 2 weeks). Appendix \ref{app:emp} reports descriptive statistics for the dataset.

In the third replication, we consider a field experiment concerned with the economics of charity \parencite{list2007}. \textcite{list2007} study the effect of mail solicitation on charitable giving in the US based on a dataset of more than 50,000 prior donors (henceforth the ``charity data''), where individuals were randomly assigned into treated and control with a treatment share of approximately $2/3$. The treated group received direct mail solicitations seeking contributions with an additional matching grant. We re-evaluate the impact of the matching grant on the donation amount in US dollars. Appendix \ref{app:charity} reports descriptive statistics for the dataset.

\subsection{Scoring}

\begin{figure}[H]
    \centering
    \includegraphics[scale=1.1]{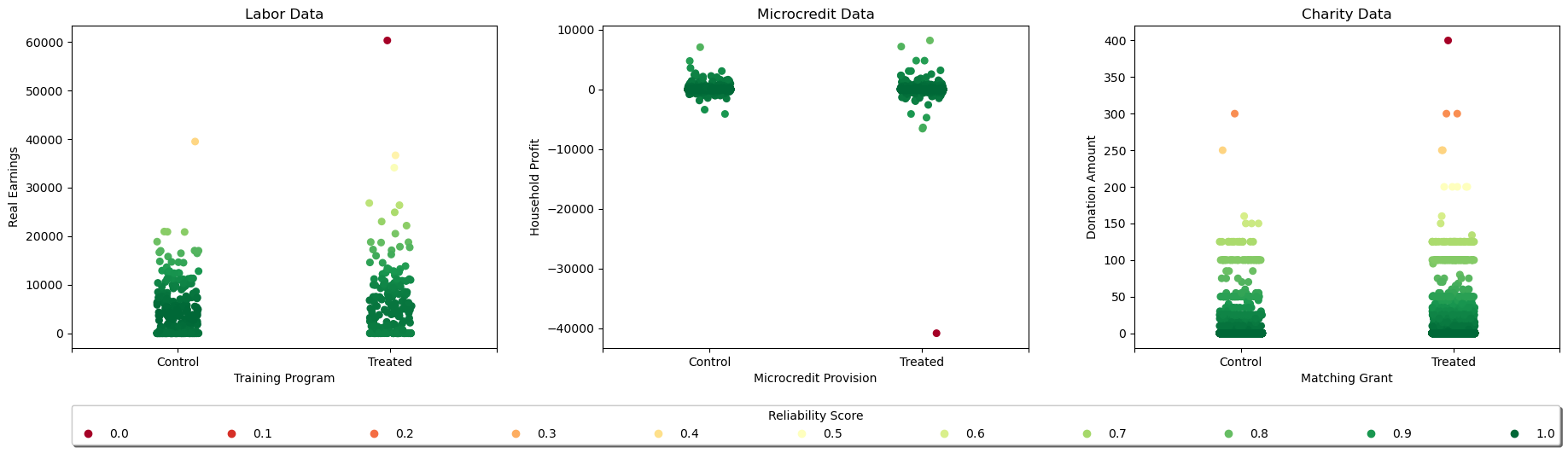}
    \caption{Scoring - Comparison for Labor, Microcredit and Charity Data}
    \label{fig:all_scores}
\end{figure}

We begin our replications by scoring data point reliability and plotting the results in Figure \ref{fig:all_scores}. For the labor data, SFR suggests that there are four data points with low reliability, representing individuals with atypically high earnings above 30,000 dollars.\footnote{For a discussion of outliers in the non-experimental version of the dataset, see \textcite{canavire2021}.} For the microcredit data, SFR reveals that there is one data point that is highly unreliable, with very high leverage, representing a single household with an extremely large loss. For the charity data, SFR suggests that the dataset contains several unreliable data points with moderate leverage, and that the unreliable data is partly balanced between treated and control. These observations represent unusually high donations compared to the rest of the sample.

In reviewing Figure \ref{fig:all_scores}, we note that low reliability scores do not automatically imply ``bad data''. Instead, the scores and plots above may motivate researchers to inspect the data more carefully and assess whether the indicated observations are corrupted by measurement error, are related to a rare or unexpected condition, or otherwise are outliers that may distort what is represented by the estimated ``average'' effect of the treatment.

\subsection{Annealing}

We continue our replications by annealing the samples and plotting the re-estimated treatment effect across the sequential removal of the most unreliable data points, along with a bootstrapped $95\%$ confidence interval for inference. 

\begin{figure}[H]
    \centering
    \includegraphics[scale=0.7]{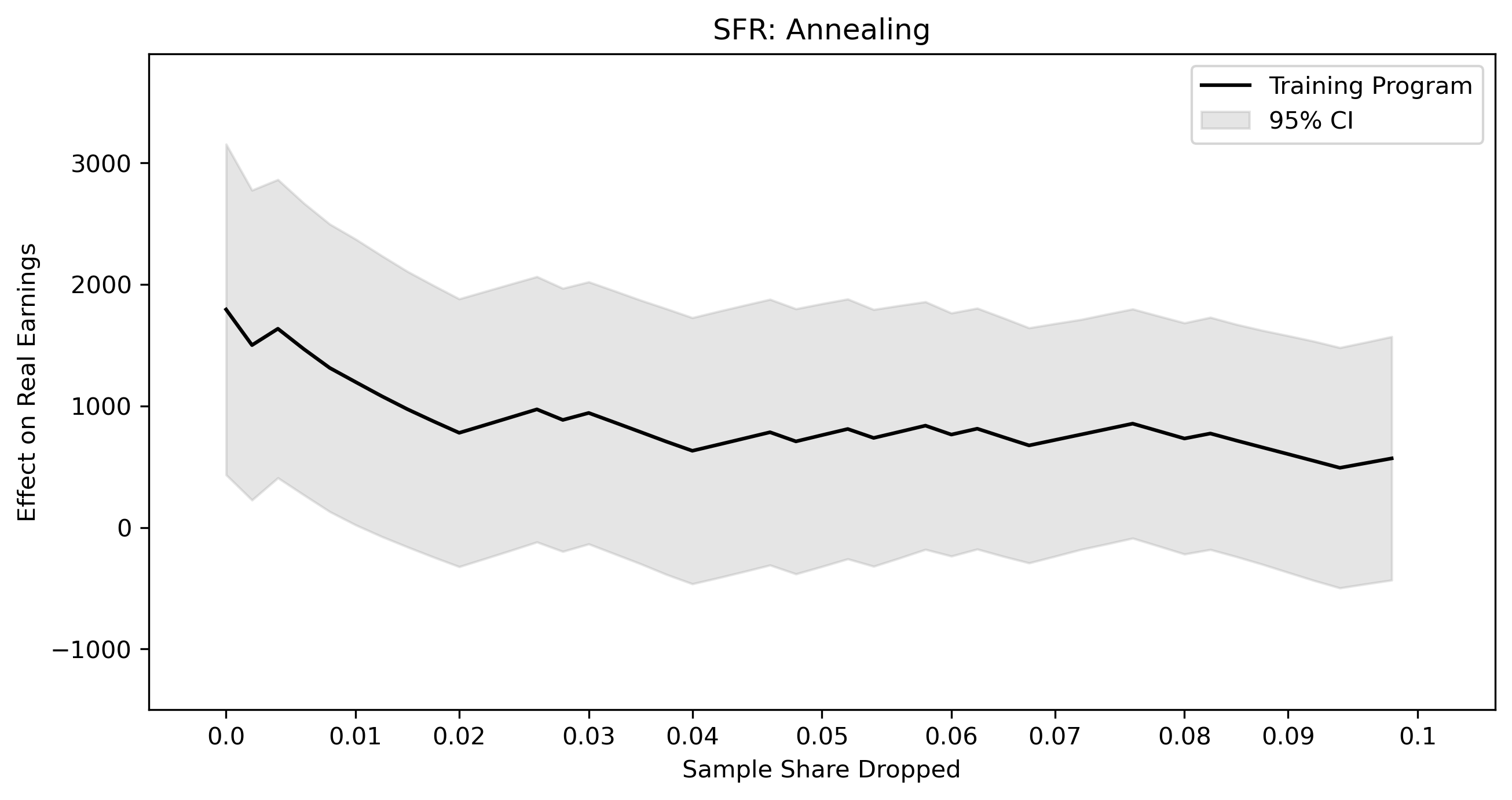}
    \caption{Annealing - Labor Data}
    \label{fig:training_annealing}
\end{figure}

In Figure \ref{fig:training_annealing}, we observe that the estimated effect of job training on earnings gradually decreases from the initial estimate of almost 2,000 dollars to around 1,000 dollars after $2\%$ of the most unreliable points are removed from the sample. The annealed effect becomes statistically indifferent from zero after about $1\%$ of the most unreliable points are removed from the sample. As such, we believe that the annealing graph enhances the substantive conclusions one can draw from the study based on the reliability of the sample.

\begin{figure}[H]
    \centering
    \includegraphics[scale=0.7]{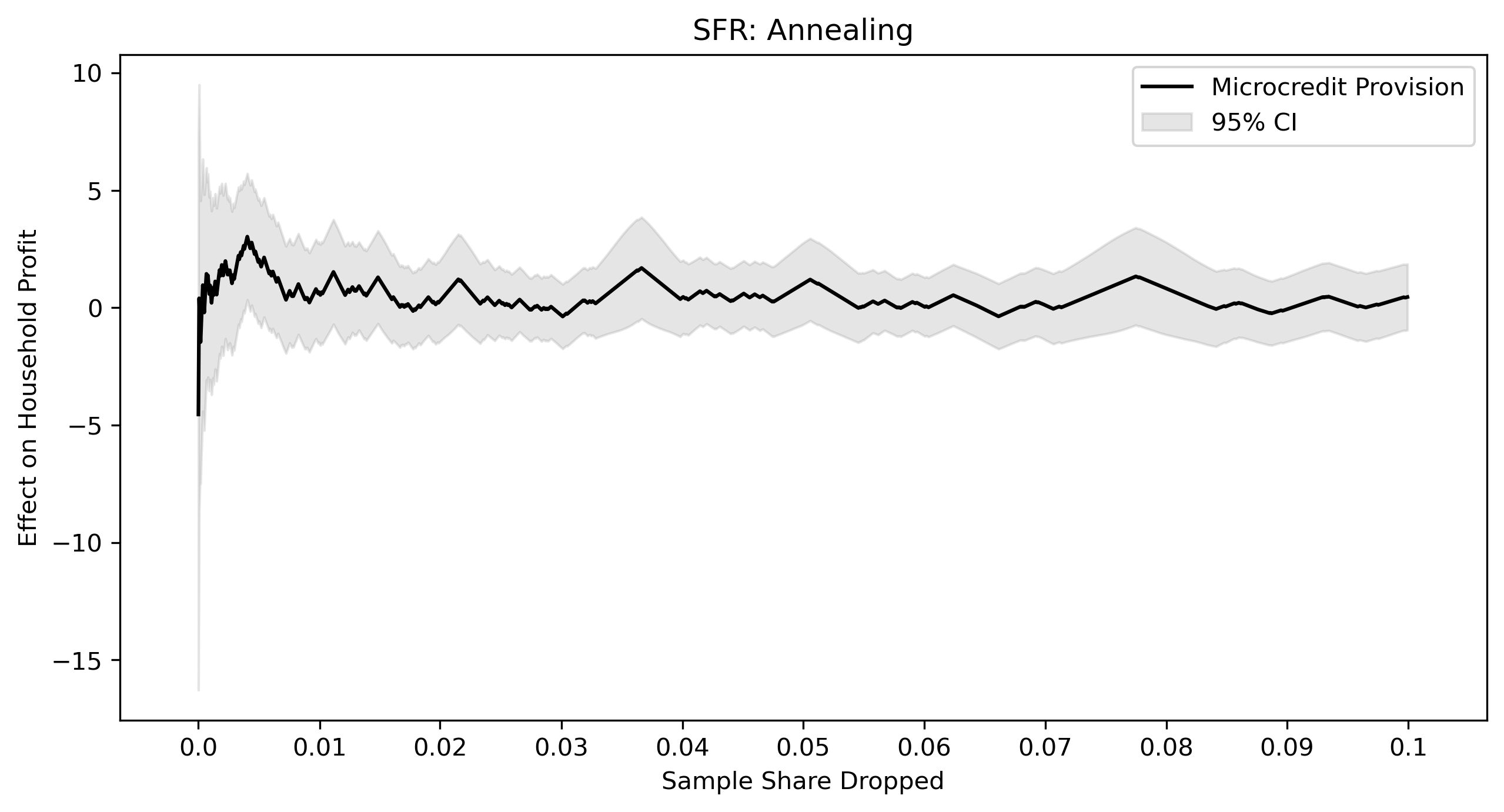}
    \caption{Annealing - Microcredit Data}
    \label{fig:microcredit_annealing}
\end{figure}

In Figure \ref{fig:microcredit_annealing}, we observe that the effect of the provision of microcredit on profits changes rapidly and inconsistently during the annealing of the first half of $1\%$ of the sample. After dropping around $1\%$ of the sample, the effect changes from negative to positive and then stabilizes around zero. We also observe instability in the estimation of the treatment effect, represented by the wide confidence interval in the region with the lowest reliability scores. Given that overall the replication does not find evidence for an effect (as the original study did not), it might appear that SFR does not change the substantive conclusions of the study. However, SFR does allow one to conclude with higher confidence that the ATE for the vast majority of the sample is in fact close to zero.

\begin{figure}[H]
    \centering
    \includegraphics[scale=0.7]{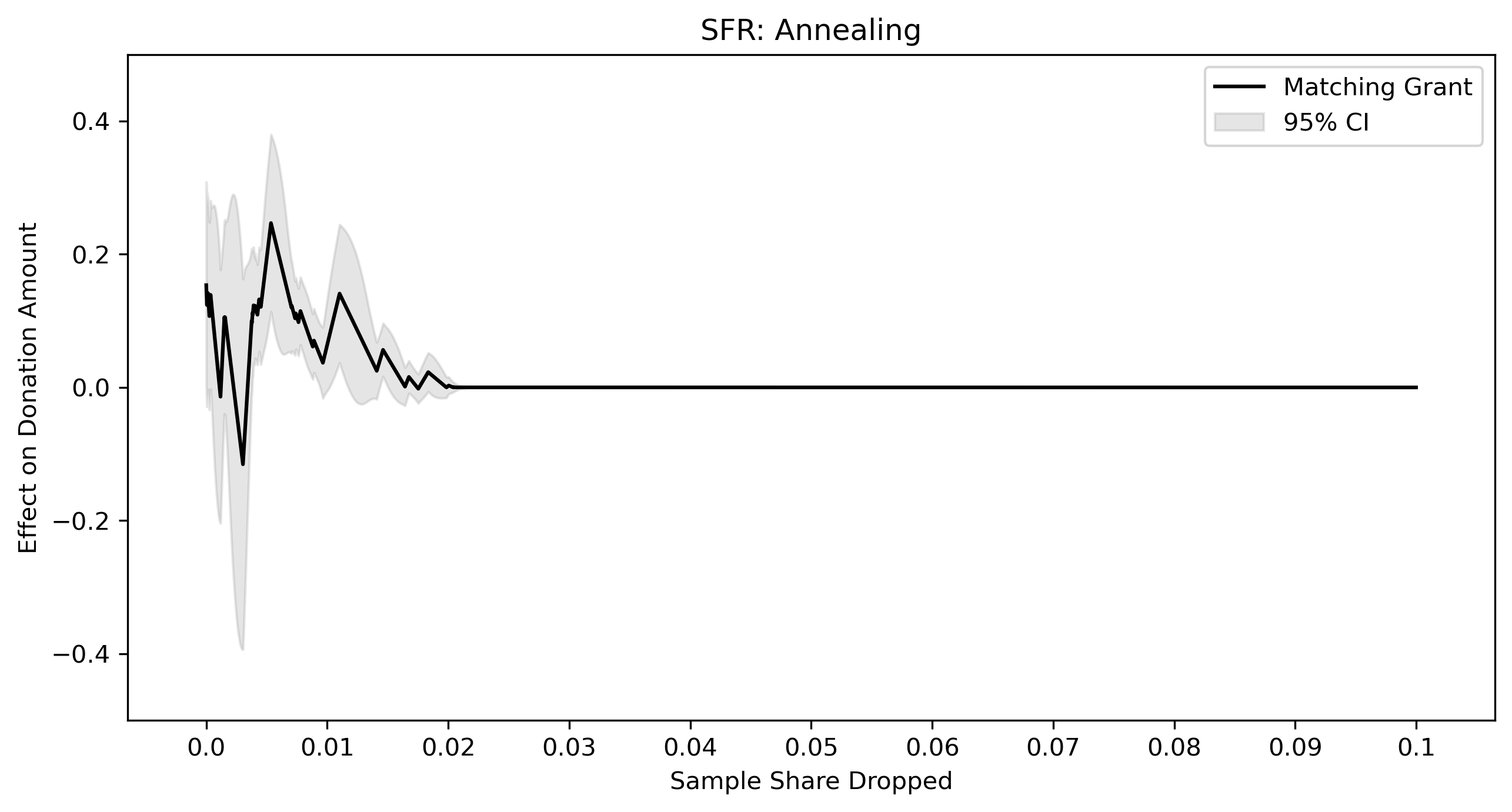}
    \caption{Annealing - Charity Data}
    \label{fig:charity_annealing}
\end{figure}

In Figure \ref{fig:charity_annealing}, we observe a clear pattern in the treatment effect of matching grants on donations. Dropping the first $2\%$ of the sample with the lowest reliability scores changes the point estimate of the ATE from a positive, to a negative, and back to a positive result (albeit with broad confidence intervals that appear to come in and out of statistical significance). Annealing the first $2\%$ stabilizes the effect close to zero, along with a tight bootstrapped confidence band. Together with scoring, annealing reveals that the average effect seems to be driven by a few donations that are exceptionally large. Moreover, there appears to be no evidence of an effect for over $98\%$ of the sample. Instead, alternative explanations might exist to shed light on the results that were annealed from the top $2\%$ of unreliable observations in the sample.\footnote{However, note the implicit assumption of effect homogeneity that is imposed by the linear model specification.} In turn, researchers might be motivated to test \parencite{crump2008} and investigate treatment effect heterogeneity in a systematic way \parencites[see e.g.,][]{athey2019}[or][for such approaches]{kunzel2019}[as well as][for such heterogeneity analysis in the context of charitable giving.]{cagala2021}.\footnote{Note that \textcite{list2007} also conduct additional, more refined analyses going beyond the difference-in-means evaluation for the aggregate effect and provide results on conditional analyses as well as effect heterogeneity analyses.}

In reviewing Figures \ref{fig:training_annealing}, \ref{fig:microcredit_annealing}, and \ref{fig:charity_annealing}, we note that restricting the sample via annealing results in the estimation of an ATE for different sub-populations, as stratified by sorted reliability scores. In other words, the annealing procedure translates the ATE to a type of a \textit{conditional} average treatment effect, i.e. CATE \parencites{abrevaya2015}{jacob2021}. We therefore recommend researchers to examine the covariate distribution between removed and remaining sub-samples to assess whether underlying sub-populations differ in practice \parencites(see for example)(for a discussion of appropriate covariate balance measures)[][or]{iacus2011}{diamond2013}.

\subsection{Fitting}
Finally, we complete our replications by executing an automatic analysis that estimates treatment effects based on reliability scores via a weighted regression with 1,000 bootstrap replications for inference. We summarize the results from the weighted regressions in Table \ref{tab:comparison_fit}, and comparisons to Huber and RANSAC estimations are provided in Appendix \ref{app:ate}. 

\begin{table}[H]
\small
\centering
\caption{Fitting - Comparison for Labor, Microcredit and Charity Data}
\label{tab:comparison_fit}
\begin{tabular}{lrrrrrr}
\toprule
\multicolumn{7}{c}{\textit{\textbf{The Effect of Training Program on Real Earnings}}}\\
\midrule
{} &     Coef. &  Std.Err. &      t &  P>|t| &     [0.025 &    0.975] \\
\midrule
OLS & 1794.3431 &  632.8536 &  2.8353 & 0.0048 & 550.5749 & 3038.1113 \\
SFR & 1206.1924 &  563.4645 &  2.1407 & 0.0328 &  98.7967 & 2313.5880 \\
\midrule
\multicolumn{7}{c}{\textit{\textbf{The Effect of Microcredit Provision on Household Profit}}}\\
\midrule
{} &   Coef. &  Std.Err. &       t &  P>|t| &   [0.025 &  0.975] \\
\midrule
OLS &   -4.5491 &    5.8789 & -0.7738 & 0.4391 & -16.0724 &    6.9741 \\
SFR &    0.5084 &    2.2366 &  0.2273 & 0.8202 &  -3.8755 &    4.8923 \\
\midrule
 \multicolumn{7}{c}{\textit{\textbf{The Effect of Matching Grants on Donation Amount}}}\\
\midrule
{} &  Coef. &  Std.Err. &       t &  P>|t| &  [0.025 &  0.975] \\
\midrule
OLS &    0.1536 &    0.0826 &  1.8605 & 0.0628 &  -0.0082 &    0.3154 \\
SFR &    0.1039 &    0.0409 &  2.5417 & 0.0110 &   0.0238 &    0.1840 \\
\bottomrule
\end{tabular}
\end{table}

In the case of the evaluation of the job training program, an SFR regression weighted by reliability scores yields a point estimate that is substantially lower than an OLS estimate, with a similar statistical precision based on the bootstrap inference. SFR results for the microcredit study also suggest an effect that is markedly different from the original OLS estimate, although the estimate is noisy and we do not find evidence for an effect that is different from zero. Lastly, SFR estimation of the effect of charitable giving yields a smaller, but more precise estimate of the ATE, compared to the original estimate.

\section{Conclusion}\label{sec:conclusion}

In this paper, we propose the concept of \textit{sample fit} as a complement to the widely used notion of model fit and argue for a more data-focused perspective in econometric models. To assess sample fit, we combine ideas from classical robust statistics and modern machine learning to develop \textit{Sample Fit Reliability} -- a computational re-sampling approach to estimate the reliability of the data, check the sensitivity of the results, and improve the robustness of the analysis. SFR entails three aspects: \textit{Scoring}, to estimate reliability scores for every data point in a sample that reflect the average estimation loss over sub-samples; \textit{Annealing}, to test the sensitivity of estimation results to the sequential removal of the most unreliable observations from the sample; and \textit{Fitting}, to estimate a weighted regression that down-weights the unreliable observations in the sample. We provide a software implementation of SFR in a Python library called \href{https://pypi.org/project/samplefit/}{samplefit} available on \textsf{PyPI} to facilitate its usage for applied empirical research.\footnote{Data and replication files can be found at: \url{https://github.com/okasag/samplefit/tree/main/replication}.}

We illustrate the application of SFR using a real-world data example; test its performance based on synthetic data in simulations; and apply SFR in replications of renowned field experiments in economics. In general, we find that SFR methods improve understanding of sample fit and reveal valuable insights about the reliability of results. As such, we join a stream of research that aims to improve sensitivity analysis, model interpretation, and other forms of robust generalization by explicitly tackling data issues to increase credibility of empirical studies.

Given a substantial amount of data work (i.e., cleaning, filtering, imputing, scaling, etc.) involved in empirical research, we aspire to shift the attention in applied econometrics to the importance of data reliability, much as the “credibility revolution” \parencite{angrist2010} shifted attention to the importance of research design. With the development of SFR, we provide one approach for assessing sample fit, but there also are complementary approaches, such as recently proposed methods by \textcite{broderick2020}, \textcite{kuschnig2021} and \textcite{moitra2022} that challenge results in an adversarial manner.

Finally, it would be valuable to extend the SFR framework to non-parametric models, including highly flexible machine learning models. We also see potential for future research that attempts to simultaneously optimize model fit, \textit{together with} sample fit, for better prediction models, better analysis of effect heterogeneity, and a better understanding of algorithmic fairness.

\pagebreak

\printbibliography

\pagebreak

\appendix
\section{Descriptive Statistics}\label{app:desc}
Below, we summarize the descriptive statistics for the non-synthetic datasets used for the illustration and replications in the main text of the manuscript.

\subsection{Boston Housing Data}\label{app:boston}

\begin{table}[H]
\centering
\footnotesize
\caption{Descriptive Statistics for the Boston Housing Data}
\begin{tabular}{lrrrrrrrr}
\toprule
{} &  count &   mean &   std &   min &   25\% &    50\% &    75\% &    max \\
\midrule
Per Capita Crime Rate  &  506 &   3.61 &  8.60 &  0.01 &  0.08 &   0.26 &   3.68 &  88.98 \\
\% Lower Status of the Population &  506 &  12.65 &  7.14 &  1.73 &  6.95 &  11.36 &  16.96 &  37.97 \\
\bottomrule
\end{tabular}
\end{table}

\begin{figure}[H]
    \centering
    \includegraphics[scale=0.65]{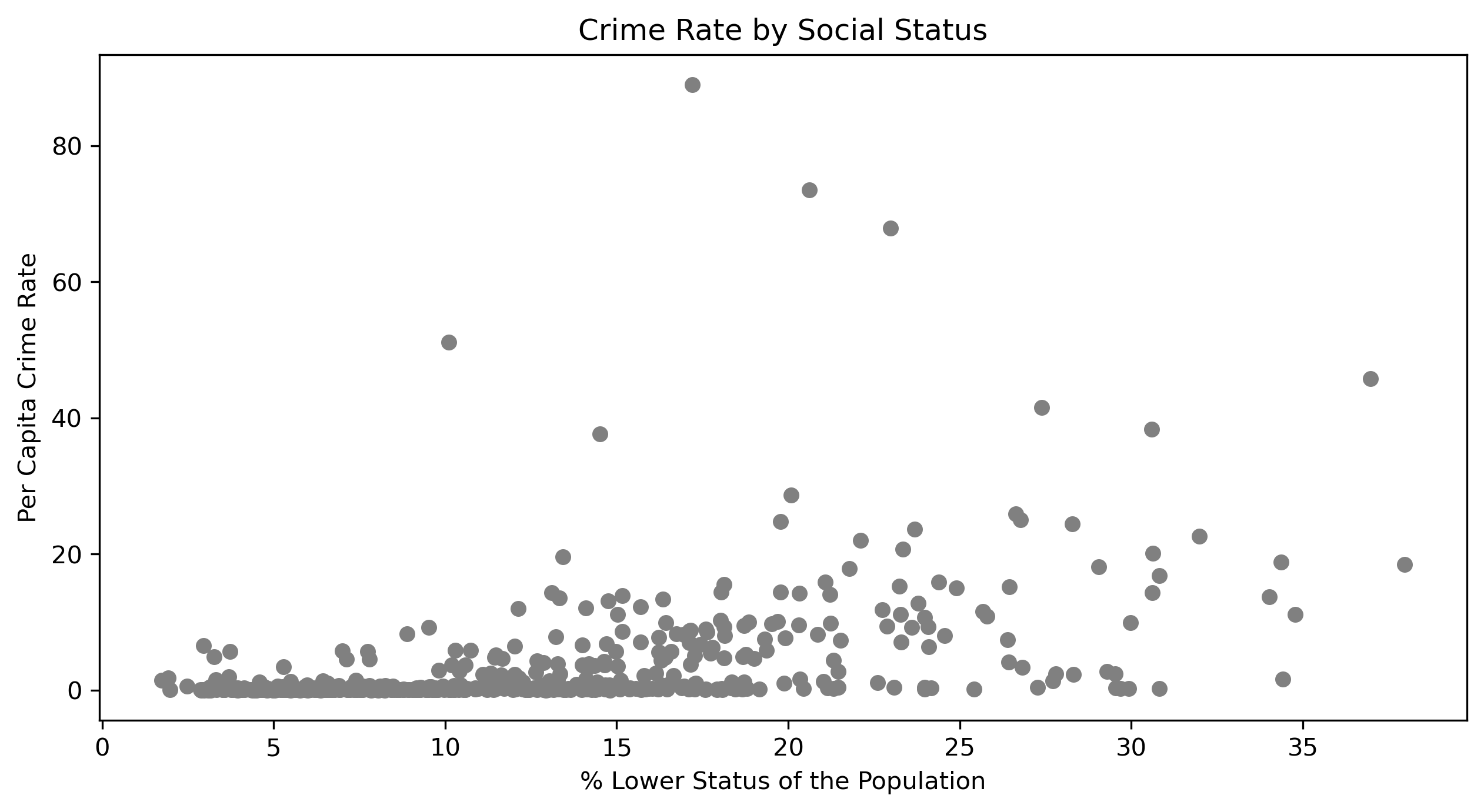}
    \caption{Scatterplot of the Boston Housing Data}
    \label{fig:scatter_boston}
\end{figure}

\subsection{Labor Data}\label{app:training}

\begin{table}[H]
\centering
\footnotesize
\caption{Descriptive Statistics for the Labor Data}
\begin{tabular}{lrrrrrrrr}
\toprule
{} &    count &      mean &       std &    min &    25\% &       50\% &       75\% &        max \\
\midrule
Real Earnings      & 445 & 5300.7700 & 6631.4900 & 0.0000 & 0.0000 & 3701.8100 & 8124.7200 & 60307.9000 \\
Training Program & 445 &    0.4200 &    0.4900 & 0.0000 & 0.0000 &    0.0000 &    1.0000 &     1.0000 \\
\bottomrule
\end{tabular}
\end{table}

\begin{figure}[H]
    \centering
    \includegraphics[scale=0.7]{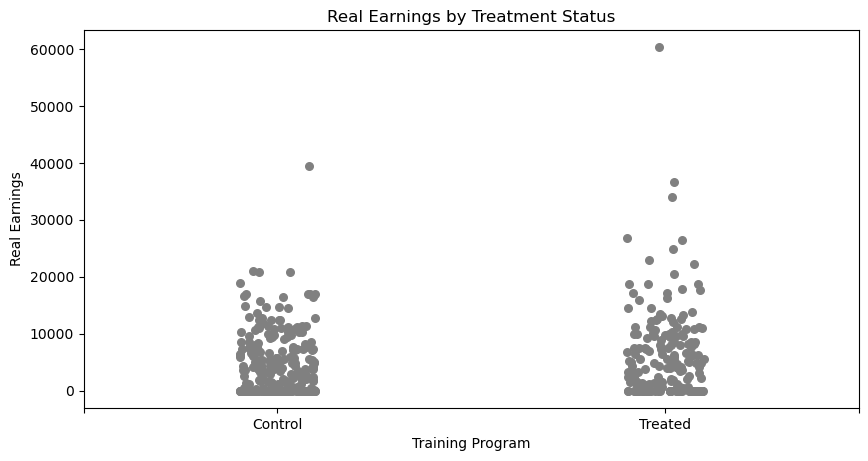}
    \caption{Scatterplot of the Labor Data}
    \label{fig:scatter_earnings}
\end{figure}

\pagebreak

\subsection{Microcredit Data}\label{app:emp}

\begin{table}[H]
\centering
\footnotesize
\caption{Descriptive Statistics for the Microcredit Data}
\begin{tabular}{lrrrrrrrr}
\toprule
{} &      count &    mean &      std &         min &    25\% &    50\% &    75\% &       max \\
\midrule
Household Profit    & 16560 & 12.1100 & 378.2600 & -40854.4200 & 0.0000 & 0.0000 & 0.0000 & 8211.9400 \\
Microcredit Provision & 16560 &  0.5000 &   0.5000 &      0.0000 & 0.0000 & 0.0000 & 1.0000 &    1.0000 \\
\bottomrule
\end{tabular}
\end{table}

\begin{figure}[H]
    \centering
    \includegraphics[scale=0.7]{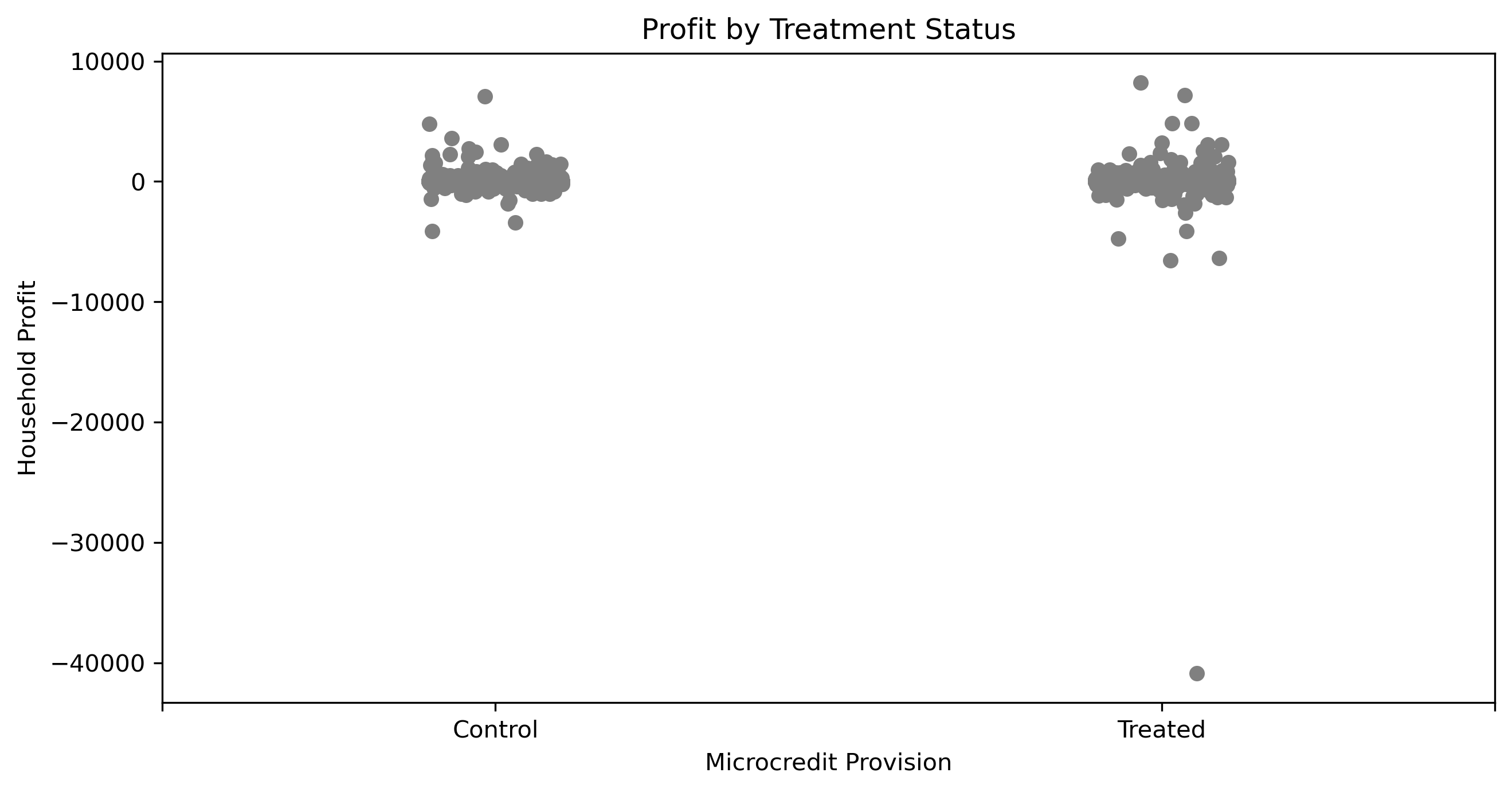}
    \caption{Scatterplot of the Microcredit Data}\label{microcredit}
\end{figure}

\subsection{Charity Data}\label{app:charity}

\begin{table}[H]
\centering
\footnotesize
\caption{Descriptive Statistics for the Charity Data}
\begin{tabular}{lrrrrrrrr}
\toprule
{} &      count &   mean &    std &    min &    25\% &    50\% &    75\% &      max \\
\midrule
Donation Amount    & 50083 & 0.9200 & 8.7100 & 0.0000 & 0.0000 & 0.0000 & 0.0000 & 400.0000 \\
Matching Grant & 50083 & 0.6700 & 0.4700 & 0.0000 & 0.0000 & 1.0000 & 1.0000 &   1.0000 \\
\bottomrule
\end{tabular}
\end{table}

\begin{figure}[H]
    \centering
    \includegraphics[scale=0.7]{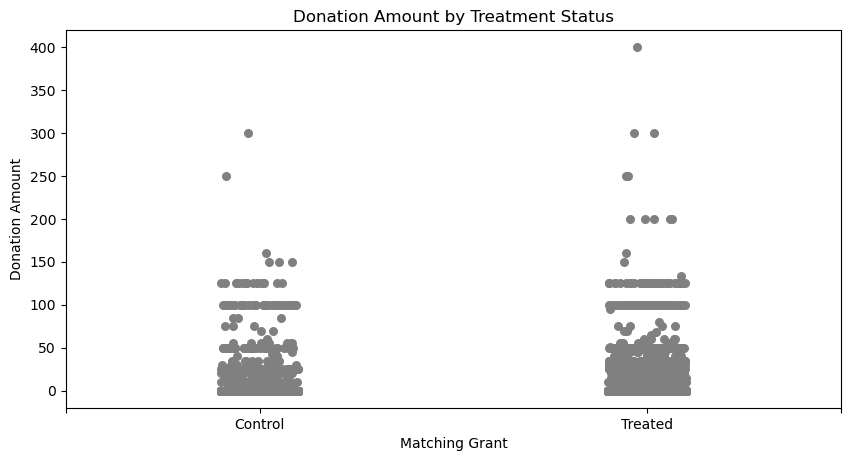}
    \caption{Scatterplot of the Charity Data}
    \label{fig:scatter_amount}
\end{figure}

\section{Supplementary Results}\label{app:results}

\subsection{Simulation Study}\label{app:sim}
Herein, we present the full set of results of the simulation study. Tables \ref{apptab:s1}, \ref{apptab:s2}, \ref{apptab:s3} and \ref{apptab:s4} summarize the results for all four simulation scenarios, namely the scenario with no outliers, scenario with randomly distributed outliers, scenario with densely distributed outliers with leverage and scenario with clusters of such densely distributed outliers with leverage, respectively. We repeat the simulation exercise for varying sample sizes $N = \{100, 500, 1000\}$ and varying outlier shares $\alpha = \{1\%, 2.5\%, 5\%\}$. We estimate the unknown coefficient $\beta$ via OLS, Huber, RANSAC and the proposed SFR estimator for a total of $R = 1000$ simulation replications. We then evaluate the performance with respect to the effect (slope) estimation based on the mean squared error, mean absolute bias, standard deviation and the Jarque Bera statistic.

\subsubsection{Scenario 1: No Outliers}\label{app:s1}

\begin{table}[!ht]
\centering
\caption{Scenario 1 - Simulation Results for Effect Estimation}
\label{apptab:s1}
\begin{tabular}{lrrr}
\toprule
\textit{\textbf{Sample Size}} &  N=100 & N=500 &  N=1000 \\
\midrule
\textit{\textbf{Outlier Share}} & 0\% &  0\% &  0\% \\
\midrule
& \multicolumn{3}{c}{\textit{\textbf{Mean Squared Error}}}\\
\midrule
OLS    &   0.0044 &   0.0009 &    0.0005 \\
HUBER  &   0.0046 &   0.0010 &    0.0005 \\
RANSAC &   0.0315 &   0.0227 &    0.0188 \\
SFR    &   0.0068 &   0.0013 &    0.0006 \\
\midrule
& \multicolumn{3}{c}{\textit{\textbf{Mean Absolute Bias}}}\\
\midrule
OLS    &   0.0534 &   0.0242 &    0.0169 \\
HUBER  &   0.0541 &   0.0248 &    0.0173 \\
RANSAC &   0.1420 &   0.1196 &    0.1090 \\
SFR    &   0.0654 &   0.0296 &    0.0201 \\
\midrule
& \multicolumn{3}{c}{\textit{\textbf{Standard Deviation}}}\\
\midrule
OLS    &   0.0667 &   0.0305 &    0.0213 \\
HUBER  &   0.0680 &   0.0311 &    0.0218 \\
RANSAC &   0.1774 &   0.1506 &    0.1371 \\
SFR    &   0.0827 &   0.0367 &    0.0253 \\
\midrule
& \multicolumn{3}{c}{\textit{\textbf{Jarque Bera Statistic}}}\\
\midrule
OLS    &   0.7445 &   0.6749 &    0.8285 \\
HUBER  &   0.3017 &   0.5732 &    0.9577 \\
RANSAC &   0.8721 &   0.6620 &    0.9553 \\
SFR    &   0.1335 &   0.8992 &    0.6744 \\
\bottomrule
\multicolumn{4}{l}{\footnotesize \textit{Note:} $p$-values for Jarque Bera statistic reported.}
\end{tabular}
\end{table}

\begin{comment}
\begin{figure}[H]
    \centering
    \includegraphics[scale=0.55]{sim1_weights.png}
    \caption{Weighting Schemes - Scenario 1: No Outliers}
    \label{fig:sim1_weights}
\end{figure}
\end{comment}

\pagebreak

\subsubsection{Scenario 2: Randomly Distributed Outliers}\label{app:s2}

\begin{table}[!ht]
\centering
\caption{Scenario 2 - Simulation Results for Effect Estimation}
\label{apptab:s2}
\begin{tabular}{lrrrrrrrrr}
\toprule
\textit{\textbf{Sample Size}} &  \multicolumn{3}{c}{N=100} & \multicolumn{3}{c}{N=500} &  \multicolumn{3}{c}{N=1000} \\
\midrule
\textit{\textbf{Outlier Share}} & 1\% &  2.5\% &  5\% &  1\% &  2.5\% &  5\% &  1\% &  2.5\% &  5\% \\
\midrule
& \multicolumn{9}{c}{\textit{\textbf{Mean Squared Error}}}\\
\midrule
OLS    &    0.0090 &     0.0169 &    0.0473 &    0.0032 &     0.0102 &    0.0321 &     0.0023 &      0.0101 &     0.0313 \\
HUBER  &    0.0052 &     0.0059 &    0.0111 &    0.0012 &     0.0019 &    0.0050 &     0.0006 &      0.0016 &     0.0046 \\
RANSAC &    0.0328 &     0.0305 &    0.0318 &    0.0243 &     0.0231 &    0.0258 &     0.0227 &      0.0226 &     0.0225 \\
SFR    &    0.0064 &     0.0063 &    0.0069 &    0.0011 &     0.0014 &    0.0032 &     0.0005 &      0.0012 &     0.0036 \\
\midrule
& \multicolumn{9}{c}{\textit{\textbf{Mean Absolute Bias}}}\\
\midrule
OLS    &    0.0725 &     0.0979 &    0.1788 &    0.0457 &     0.0893 &    0.1687 &     0.0403 &      0.0932 &     0.1714 \\
HUBER  &    0.0570 &     0.0610 &    0.0843 &    0.0273 &     0.0359 &    0.0625 &     0.0202 &      0.0338 &     0.0630 \\
RANSAC &    0.1420 &     0.1374 &    0.1393 &    0.1243 &     0.1199 &    0.1274 &     0.1222 &      0.1202 &     0.1176 \\
SFR    &    0.0633 &     0.0631 &    0.0661 &    0.0261 &     0.0296 &    0.0478 &     0.0187 &      0.0282 &     0.0536 \\
\midrule
& \multicolumn{9}{c}{\textit{\textbf{Standard Deviation}}}\\
\midrule
OLS    &    0.0869 &     0.1073 &    0.1337 &    0.0412 &     0.0490 &    0.0600 &     0.0294 &      0.0376 &     0.0440 \\
HUBER  &    0.0709 &     0.0730 &    0.0819 &    0.0318 &     0.0328 &    0.0362 &     0.0220 &      0.0243 &     0.0257 \\
RANSAC &    0.1810 &     0.1746 &    0.1783 &    0.1557 &     0.1519 &    0.1607 &     0.1507 &      0.1504 &     0.1495 \\
SFR    &    0.0797 &     0.0794 &    0.0773 &    0.0325 &     0.0323 &    0.0353 &     0.0218 &      0.0244 &     0.0273 \\
\midrule
& \multicolumn{9}{c}{\textit{\textbf{Jarque Bera Statistic}}}\\
\midrule
OLS    &    0.0000 &     0.0000 &    0.0000 &    0.0000 &     0.0000 &    0.0000 &     0.0024 &      0.0010 &     0.0001 \\
HUBER  &    0.1326 &     0.1252 &    0.0062 &    0.6158 &     0.8013 &    0.1715 &     0.2982 &      0.6733 &     0.4350 \\
RANSAC &    0.0967 &     0.0278 &    0.0001 &    0.4091 &     0.2634 &    0.0201 &     0.0328 &      0.5448 &     0.4082 \\
SFR    &    0.9362 &     0.4214 &    0.0989 &    0.1580 &     0.6732 &    0.0420 &     0.0100 &      0.4236 &     0.1535 \\
\bottomrule
\multicolumn{10}{l}{\footnotesize \textit{Note:} $p$-values for Jarque Bera statistic reported.}
\end{tabular}
\end{table}

\begin{comment}
\begin{figure}[H]
    \centering
    \includegraphics[scale=0.55]{sim2_weights.png}
    \caption{Weighting Schemes - Scenario 2: Randomly Distributed Outliers}
    \label{fig:sim2_weights}
\end{figure}
\end{comment}

\pagebreak

\subsubsection{Scenario 3: Densely Distributed Outliers with Leverage}\label{app:s3}

\begin{table}[!ht]
\centering
\caption{Scenario 3 - Simulation Results for Effect Estimation}
\label{apptab:s3}
\begin{tabular}{lrrrrrrrrr}
\toprule
\textit{\textbf{Sample Size}} &  \multicolumn{3}{c}{N=100} & \multicolumn{3}{c}{N=500} &  \multicolumn{3}{c}{N=1000} \\
\midrule
\textit{\textbf{Outlier Share}} & 1\% &  2.5\% &  5\% &  1\% &  2.5\% &  5\% &  1\% &  2.5\% &  5\% \\
\midrule
& \multicolumn{9}{c}{\textit{\textbf{Mean Squared Error}}}\\
\midrule
OLS    &    0.0089 &     0.0185 &    0.0753 &    0.0047 &     0.0197 &    0.0684 &     0.0041 &      0.0208 &     0.0674 \\
HUBER  &    0.0054 &     0.0069 &    0.0211 &    0.0015 &     0.0040 &    0.0154 &     0.0009 &      0.0038 &     0.0151 \\
RANSAC &    0.0340 &     0.0309 &    0.0314 &    0.0240 &     0.0226 &    0.0234 &     0.0226 &      0.0219 &     0.0209 \\
SFR    &    0.0060 &     0.0061 &    0.0101 &    0.0012 &     0.0022 &    0.0082 &     0.0007 &      0.0021 &     0.0091 \\
\midrule
& \multicolumn{9}{c}{\textit{\textbf{Mean Absolute Bias}}}\\
\midrule
OLS    &    0.0750 &     0.1182 &    0.2623 &    0.0619 &     0.1363 &    0.2591 &     0.0600 &      0.1422 &     0.2585 \\
HUBER  &    0.0582 &     0.0656 &    0.1264 &    0.0312 &     0.0554 &    0.1195 &     0.0254 &      0.0574 &     0.1207 \\
RANSAC &    0.1446 &     0.1391 &    0.1372 &    0.1238 &     0.1191 &    0.1215 &     0.1220 &      0.1178 &     0.1132 \\
SFR    &    0.0616 &     0.0618 &    0.0813 &    0.0270 &     0.0388 &    0.0832 &     0.0209 &      0.0404 &     0.0918 \\
\midrule
& \multicolumn{9}{c}{\textit{\textbf{Standard Deviation}}}\\
\midrule
OLS    &    0.0715 &     0.0720 &    0.0811 &    0.0316 &     0.0332 &    0.0355 &     0.0216 &      0.0249 &     0.0243 \\
HUBER  &    0.0698 &     0.0699 &    0.0764 &    0.0311 &     0.0319 &    0.0336 &     0.0215 &      0.0235 &     0.0236 \\
RANSAC &    0.1843 &     0.1757 &    0.1771 &    0.1548 &     0.1502 &    0.1529 &     0.1504 &      0.1478 &     0.1447 \\
SFR    &    0.0768 &     0.0766 &    0.0772 &    0.0322 &     0.0330 &    0.0354 &     0.0221 &      0.0247 &     0.0258 \\
\midrule
& \multicolumn{9}{c}{\textit{\textbf{Jarque Bera Statistic}}}\\
\midrule
OLS    &    0.0016 &     0.0077 &    0.7897 &    0.0830 &     0.6347 &    0.9059 &     0.5048 &      0.5940 &     0.4811 \\
HUBER  &    0.2798 &     0.0153 &    0.3334 &    0.7088 &     0.6461 &    0.5829 &     0.1747 &      0.6779 &     0.2055 \\
RANSAC &    0.0839 &     0.1732 &    0.0000 &    0.2323 &     0.8904 &    0.0591 &     0.0361 &      0.5644 &     0.0137 \\
SFR    &    0.8841 &     0.0286 &    0.0443 &    0.0811 &     0.8855 &    0.8651 &     0.0531 &      0.7285 &     0.6346 \\
\bottomrule
\multicolumn{10}{l}{\footnotesize \textit{Note:} $p$-values for Jarque Bera statistic reported.}
\end{tabular}
\end{table}

\begin{comment}
\begin{figure}[H]
    \centering
    \includegraphics[scale=0.55]{sim3_weights.png}
    \caption{Weighting Schemes - Scenario 3: Asymmetrically Leveraged Outliers}
    \label{fig:sim3_weights}
\end{figure}
\end{comment}

\pagebreak

\subsubsection{Scenario 4: Clusters of Densely Distributed Outliers with Leverage}\label{app:s4}

\begin{table}[!ht]
\centering
\caption{Scenario 4 - Simulation Results for Effect Estimation}
\label{apptab:s4}
\begin{tabular}{lrrrrrrrrr}
\toprule
\textit{\textbf{Sample Size}} &  \multicolumn{3}{c}{N=100} & \multicolumn{3}{c}{N=500} &  \multicolumn{3}{c}{N=1000} \\
\midrule
\textit{\textbf{Outlier Share}} & 1\% &  2.5\% &  5\% &  1\% &  2.5\% &  5\% &  1\% &  2.5\% &  5\% \\
\midrule
& \multicolumn{9}{c}{\textit{\textbf{Mean Squared Error}}}\\
\midrule
OLS    &    0.0088 &     0.0192 &    0.0808 &    0.0046 &     0.0205 &    0.0733 &     0.0041 &      0.0217 &     0.0734 \\
HUBER  &    0.0054 &     0.0069 &    0.0210 &    0.0015 &     0.0040 &    0.0153 &     0.0009 &      0.0038 &     0.0151 \\
RANSAC &    0.0335 &     0.0294 &    0.0297 &    0.0244 &     0.0222 &    0.0241 &     0.0226 &      0.0215 &     0.0212 \\
SFR    &    0.0059 &     0.0060 &    0.0098 &    0.0012 &     0.0022 &    0.0076 &     0.0007 &      0.0021 &     0.0085 \\
\midrule
& \multicolumn{9}{c}{\textit{\textbf{Mean Absolute Bias}}}\\
\midrule
OLS    &    0.0749 &     0.1206 &    0.2725 &    0.0612 &     0.1394 &    0.2683 &     0.0603 &      0.1453 &     0.2696 \\
HUBER  &    0.0582 &     0.0663 &    0.1257 &    0.0310 &     0.0555 &    0.1191 &     0.0253 &      0.0574 &     0.1205 \\
RANSAC &    0.1432 &     0.1352 &    0.1352 &    0.1249 &     0.1182 &    0.1227 &     0.1218 &      0.1171 &     0.1134 \\
SFR    &    0.0612 &     0.0611 &    0.0800 &    0.0268 &     0.0385 &    0.0799 &     0.0209 &      0.0396 &     0.0886 \\
\midrule
& \multicolumn{9}{c}{\textit{\textbf{Standard Deviation}}}\\
\midrule
OLS    &    0.0711 &     0.0730 &    0.0810 &    0.0309 &     0.0334 &    0.0364 &     0.0210 &      0.0243 &     0.0258 \\
HUBER  &    0.0698 &     0.0700 &    0.0767 &    0.0310 &     0.0320 &    0.0339 &     0.0214 &      0.0234 &     0.0237 \\
RANSAC &    0.1829 &     0.1713 &    0.1723 &    0.1563 &     0.1489 &    0.1552 &     0.1502 &      0.1466 &     0.1456 \\
SFR    &    0.0761 &     0.0755 &    0.0770 &    0.0322 &     0.0327 &    0.0351 &     0.0220 &      0.0245 &     0.0256 \\
\midrule
& \multicolumn{9}{c}{\textit{\textbf{Jarque Bera Statistic}}}\\
\midrule
OLS    &    0.0008 &     0.6526 &    0.0435 &    0.4527 &     0.6315 &    0.5505 &     0.5306 &      0.6379 &     0.5905 \\
HUBER  &    0.2792 &     0.4310 &    0.0075 &    0.5992 &     0.3757 &    0.3677 &     0.2130 &      0.7204 &     0.7028 \\
RANSAC &    0.0911 &     0.0626 &    0.0000 &    0.3550 &     0.7779 &    0.0012 &     0.0676 &      0.6630 &     0.0000 \\
SFR    &    0.6505 &     0.0482 &    0.1984 &    0.0337 &     0.6140 &    0.7081 &     0.0286 &      0.9678 &     0.2735 \\
\bottomrule
\multicolumn{10}{l}{\footnotesize \textit{Note:} $p$-values for Jarque Bera statistic reported.}
\end{tabular}
\end{table}

\begin{comment}
\begin{figure}[H]
    \centering
    \includegraphics[scale=0.55]{sim4_weights.png}
    \caption{Weighting Schemes - Scenario 4: Symmetrically Leveraged Outliers}
    \label{fig:sim4_weights}
\end{figure}
\end{comment}

\pagebreak

\subsection{Treatment Effects Estimation}\label{app:ate}

\begin{table}[H]
\centering
\caption{Comparison of Estimates for Average Treatment Effects}
\label{tab:comparison_fit_all}
\begin{tabular}{lrrrrrr}
\toprule
& \multicolumn{6}{c}{\textit{\textbf{The Effect of Training Program on Real Earnings}}}\\
\midrule
{} &     Coef. &  Std.Err. &      t &  P>|t| &     [0.025 &    0.975] \\
\midrule
OLS    & 1794.3431 &  632.8536 & 2.8353 & 0.0048 &   550.5749 & 3038.1113 \\
HUBER  & 1165.5045 &  491.5467 & 2.3711 & 0.0182 &   199.4514 & 2131.5577 \\
RANSAC &  193.5795 &  914.6105 & 0.2117 & 0.8325 & -1603.9350 & 1991.0940 \\
SFR    & 1206.1924 &  563.4645 & 2.1407 & 0.0328 &    98.7967 & 2313.5880 \\
\midrule
& \multicolumn{6}{c}{\textit{\textbf{The Effect of Microcredit Provision on Household Profit}}}\\
\midrule
{} &   Coef. &  Std.Err. &       t &  P>|t| &   [0.025 &  0.975] \\
\midrule
OLS    & -4.5491 &    5.8789 & -0.7738 & 0.4391 & -16.0724 &  6.9741 \\
HUBER  &  0.0000 &    0.0000 &  2.0728 & 0.0382 &   0.0000 &  0.0000 \\
RANSAC &  0.0000 &    0.0000 &     - &    - &   0.0000 &  0.0000 \\
SFR    &  0.5084 &    2.2366 &  0.2273 & 0.8202 &  -3.8755 &  4.8923 \\
\midrule
& \multicolumn{6}{c}{\textit{\textbf{The Effect of Matching Grants on Donation Amount}}}\\
\midrule
{} &  Coef. &  Std.Err. &       t &  P>|t| &  [0.025 &  0.975] \\
\midrule
OLS    & 0.1536 &    0.0826 &  1.8605 & 0.0628 & -0.0082 &  0.3154 \\
HUBER  & 0.0000 &    0.0000 & 20.4995 & 0.0000 &  0.0000 &  0.0000 \\
RANSAC & 0.0000 &    0.0000 &     - &    - &  0.0000 &  0.0000 \\
SFR    & 0.1039 &    0.0409 &  2.5417 & 0.0110 &  0.0238 &  0.1840 \\
\bottomrule
\multicolumn{7}{l}{\footnotesize \textit{Note:} SFR and RANSAC inference based on 1,000 bootstrap replications.}
\end{tabular}
\end{table}

\end{document}